\documentclass[twocolumn, pt12, twocolappendix]{aastex631}

\usepackage[T1]{fontenc}
\usepackage{graphicx}	
\usepackage{amsmath}	
\usepackage{amssymb}	
\usepackage{hyperref}
\usepackage{makecell}
\usepackage{color} 		
\usepackage{comment}    

\newcommand{\R} {\text{RMS}}

\newcommand{\radshock} {\texttt{radshock}}
\newcommand{\komrad} {\texttt{Komrad}}

\shortauthors{Samuelsson \& Ryde}
\shorttitle{Observational characteristics of RMSs in GRBs}

\begin{document}
\label{firstpage}
\title{Observational characteristics of radiation-mediated shocks in photospheric gamma-ray burst emission}


\correspondingauthor{Filip Samuelsson}
\email{filipsam@kth.se}

\author[0000-0001-7414-5884]{Filip Samuelsson}
\affiliation{Institut d'Astrophysique de Paris, Sorbonne Universit\'e and CNRS, UMR 7095, 98 bis bd Arago, F-75014 Paris, France}
\affiliation{Department of Physics, KTH Royal Institute of Technology, and The Oskar Klein Centre, SE-10691 Stockholm, Sweden}

\author[0000-0002-9769-8016]{Felix Ryde}
\affiliation{Department of Physics, KTH Royal Institute of Technology, and The Oskar Klein Centre, SE-10691 Stockholm, Sweden}




\begin{abstract}
Emission from the photosphere in gamma-ray burst (GRB) jets can be substantially affected by subphotospheric energy dissipation, which is typically caused by radiation-mediated shocks (RMSs). 
We study the observational characteristics of such emission, in particular the spectral signatures. 
Relevant shock initial conditions are estimated using a simple internal collision framework, which then serve as inputs for an RMS model that generates synthetic photospheric spectra. 
Within this framework, we find that if the free fireball acceleration starts at $r_0 \sim 10^{10}~$cm, in agreement with hydrodynamical simulations, then the typical spectrum consists of a broad, soft power-law segment with a cutoff at high energies and a hardening in X-rays. 
The synthetic spectra are generally well fitted with a standard cutoff power-law (CPL) function, as the hardening in X-rays is commonly outside the observable energy range of current detectors. The CPL-fits yield values for the low-energy index, $\alpha$, and the peak energy, $E_{\rm peak}$, that are centered around $\sim -0.8$ and $\sim 220~$keV, respectively, similar to typical observed values. 
We also identify a non-negligible parameter region for what we call ``optically shallow shocks'': shocks that do not accumulate enough scatterings to reach a steady-state spectrum before decoupling and thereby produce more complex spectra. These occur for optical depths $\tau \lesssim 55 \, u_u^{-2}$, where $u_u = \gamma_u\beta_u$ is the dimensionless specific momentum of the upstream as measured in the shock rest frame.
\end{abstract}


\section{Introduction}
In a gamma-ray burst (GRB) jet, the photon mean free path is much longer than the kinetic scales of the particles in the plasma. This means that in the regions where the jet is optically thick, sufficiently fast shocks will be mediated by radiation \citep[e.g.,][]{BlandfordPayneI1981,BlandfordPayneII1981}.
The photon spectrum in the downstream of a radiation-mediated shock (RMS) is different compared to the synchrotron spectrum emitted by high-energy particles accelerated in collisionless shocks \citep[see][for a recent review on RMSs]{LevinsonNakar2020}. 
In an RMS, the photons themselves dissipate the incoming particle kinetic energy, which, in the case of an at most mildly relativistic shock with upstream dimensionless specific momentum of $u_u \equiv \beta_u\gamma_u \lesssim~$a few, results in an extended power-law photon spectrum with a cutoff at high energy. Here, $\beta_u$ is the velocity in units $c$ of the incoming plasma as measured in the shock rest frame and $\gamma_u = (1-\beta_u^2)^{-1/2}$. Such spectra resemble those observed in the prompt phase of GRBs \citep{Lundman2018, Ito2018, Samuelsson2022}.


GRB prompt spectra are commonly fitted with a phenomenological cutoff power-law (CPL) function \citep[e.g.,][]{Yu2016, Yu2019, Poolakkil2021}. As the name suggests, the CPL function consists of a power-law segment with an exponential cutoff: $N_E \propto E^{\alpha}e^{-E/E_c}$, where $N_E$ is the specific photon number flux, $E$ is the photon energy, $\alpha$ is the spectral index, and $E_c$ is the cutoff energy. The $\alpha$-distribution found in time-resolved analysis of GRB prompt spectra with the CPL function is a smooth distribution centered at $\alpha \sim -0.8$ \citep[e.g.,][]{Yu2016}. 


The soft (i.e., low) values of $\alpha$ commonly found in catalogues of GRB observations are difficult to obtain in non-dissipative photospheric models, which generate hard observed spectra with $\alpha \gtrsim -0.5$ and higher \citep{DengZhang2014, Acuner2019}. 
However, dissipation below the photosphere can greatly alter the shape of the photon spectrum, which may not have time to relax to a thermal equilibrium before decoupling from the plasma \citep{ReesMeszaros2005}. Energy dissipation via RMSs is a natural expectation in the optically thick regions of a GRB jet. Firstly, when the jet is drilling through the star or ejected neutron star material, a high-pressure cocoon forms that keeps the jet collimated. This leads to collimation shocks extending up to, and in some cases even above, the edge of the collimating material \citep{Lazzati2009, Lopez-Camara2013, Gottlieb2018, Gottlieb2019, Gottlieb2021}. Secondly, irregularities within the jet can lead to internal shocks occurring below the photosphere. The irregularities can form either due to mixing between the jet and the surrounding material or stem from a variable central engine \citep{Bromberg2011b, Gottlieb2019}.

As a natural cause for subphotospheric dissipation, RMSs have been thoroughly discussed in the context of GRBs \citep{Eichler1994, LevinsonBromberg2008, Katz2010, Budnik2010, Bromberg2011b, Levinson2012, KerenLevinson2014, Beloborodov2017, Ito2018, Lundman2018, LundmanBeloborodov2019, LevinsonNakar2020, LundmanBeloborodov2021, Samuelsson2022}. Unfortunately, the vast difference of interaction scales between the photons and the plasma particles and the high number of photons per charged particle render simulations of RMSs that properly couple the radiation and the plasma extremely time consuming. Therefore, there existed a need for an approximate, faster method that could be used when fitting the model to data. In \citet{Samuelsson2022}, we developed such an approximation called the Kompaneets RMS approximation (KRA). The KRA allows us to accurately reproduce spectra from full-scale radiation hydrodynamic simulations using four orders of magnitude less computational time. This massive time reduction made it possible to probe a large enough parameter space such that we could perform a spectral fit using an RMS model to prompt GRB data for the first time.

In this paper, we use the KRA to investigate what typical observational signatures are expected from RMSs in GRB jets. Specifically, we focus on the observed spectrum. To estimate relevant RMS initial conditions in the context of GRBs, we employ a simple internal collision framework (the advantages and disadvantages of this approach are discussed in Section \ref{sec:Methodology}). Within this framework, we use the KRA to generate synthetic photospheric spectra, whose typical spectral characteristics can be studied. Furthermore, we compare the synthetic spectra to catalogue distributions of real observations by forward-folding the spectra through the response matrix of the \textit{Gamma-ray Burst Monitor} \citep[GBM,][]{Meegan2009} onboard the \textit{Fermi Gamma-ray Space Telescope} and subsequently fitting them with a CPL function.


The paper is organized as follows. In Section \ref{sec:KRA}, we give a short summary of the KRA. 
In Section \ref{sec:Shallow_shocks}, we show that a large parameter range results in shocks where the radiation do not have time to reach steady state, which we call ``optically shallow shocks''. In Section \ref{sec:Methodology}, we describe the methodology: we briefly outline the internal collision framework used in \ref{sec:internal_collision}, the details of which are given in Appendix \ref{App:internal_collision}, show how to go from the internal collision parameters to the KRA parameters in \ref{sec:implementation}, and describe how we obtain a CPL fit from the KRA parameters in \ref{sec:synthetic_spectra}. In Section \ref{sec:Results}, we present our results. In Section \ref{sec:Assumptions}, we list the underlying assumptions of the model and in Section \ref{sec:Discussion}, we discuss our findings. Finally, we summarize and conclude in Section \ref{sec:Summary}. We will employ the notation $Q_x = Q/10^x$ throughout the text.  



\section{The Kompaneets RMS Approximation}\label{sec:KRA}
In this section, we give a brief summary of the KRA for completeness. We refer the interested reader to \citet{Samuelsson2022} for more details. 

The KRA is an approximate but accurate method to simulate RMSs using only a tiny fraction of the computational time required by self-consistent radiation hydrodynamic simulations. One run with \komrad, the simulation code that implements the KRA, takes $\sim 2$ minutes to run on a modern laptop. A corresponding run using the full-scale radiation hydrodynamic simulation \radshock\ \citep{Lundman2018} takes several days on tens of cores. KRA is valid for mildly relativistic or slower shocks ($u_u \lesssim 3$) and in regions where radiation pressure dominates over magnetic pressure. In an internal collision framework, the former condition translates into $\Gamma_f/\Gamma_s \lesssim 36$, where $\Gamma_f$ and $\Gamma_s$ are the bulk Lorentz factors of the faster and slower blob, respectively \citep[see Appendix B in][]{Samuelsson2022}. As appropriate in the context of GRBs, the shock is assumed to be photon rich, meaning that the photon number in the downstream is dominated by advection of photons from the upstream \citep{Bromberg2011b}. The KRA can model shocks that have finished their dissipation in the optically thick regions. Specifically, we let the RMS propagate for a doubling of the radius, as this corresponds to the shock crossing time of the causally connected region. That means that we do not model shock breakout radiation, which results in more complicated dynamics \citep[e.g.,][]{LundmanBeloborodov2021}. 

As photons traverse an RMS, they experience a converging flow. The photons scatter in the velocity gradient and, on average, they gain energy in each scattering. This leads to a bulk Comptonization of the photon spectrum. As both the average relative energy gain per scattering, $\langle \Delta \epsilon / \epsilon \rangle$, and the probability of escape from the RMS are energy independent, a power-law spectrum forms. Here, $\epsilon$ is the photon energy and $\Delta \epsilon$ is the energy change in a scattering, both given in units of electron rest mass energy, $m_e c^2$. The bulk Comptonization is a Fermi-process, analogous to, e.g., the acceleration of cosmic rays in diffusive shock acceleration. In a planar parallel geometry and with a thermal upstream photon distribution, the evolution of the photon spectrum in the RMS is fully characterized by three parameters: the temperature of the photons far upstream, $\theta_u$, the incoming velocity of the upstream as measured in the shock rest frame, $\beta_u$, and the number density of photons per baryon (here taken to be protons) in the upstream, $\tilde{n} \equiv n_\gamma/n_p$. In a photon rich shock, the photon production rate is negligible and $\tilde{n}$ is roughly constant across the RMS transition region. 

The principle process described above is remarkably similar to that of thermal Comptonization of photons, which occurs if one continuously injects low-energy photons into a region of hot (nonrelativistic) electrons, and let the escape probability be energy independent \citep[e.g.,][]{RybickiLightman1979}. Thermal Comptonization is described by the Kompaneets equation, which has the major advantage of being numerically very fast to solve. This realization is the basis of the analogous and approximate description of RMSs introduced in the KRA. In the KRA, we split the, in reality, continuous RMS transition region into three discrete zones: the upstream, the RMS, and the downstream zone. In each zone, the photon distribution is evolved with the Kompaneets equation and photons are advected from the upstream to the downstream, via the RMS, using source terms. Dissipation in the RMS zone is modeled by prescribing the electrons with a high, effective electron temperature, $\theta_r$. Here and throughout the paper, the subscripts $u$, $r$, and $d$ when used for the KRA parameters refer to quantities measured in the upstream, RMS, and downstream zones, respectively. When the subscripts $u$ or $d$ are used for the physical RMS parameters, they described quantities measured in the far upstream or downstream, respectively. Furthermore, all temperatures are proper (comoving with respect to each region) and given in units of $m_ec^2$.

Apart from the effective temperature of the shock, there are two additional parameters that describe the evolution of the photon distribution in the RMS zone in the KRA. The second parameter is $R$, defined as the ratio of $\theta_r$ to the upstream temperature, $\theta_{u, {\rm K}}$, i.e., $R = \theta_r/\theta_{u, {\rm K}}$, where the subscript K stands for \komrad. The distinction is necessary as in fact $\theta_{u, {\rm K}} \neq \theta_u$. The upstream temperatures are different because in a real RMS, adiabatic compression from the upstream to the downstream results in an achromatic energy increase, which is not captured by \komrad. Hence, \komrad\ requires a higher upstream temperature in order to match the low-energy cutoff in the final spectrum (see equation \eqref{eq:theta_u_K} below). 

The final free parameter is the Compton $y$-parameter of the RMS zone, $y_r$. The Compton $y$-parameter is a measure of the average photon energy gain. In the KRA, it determines the hardness of the power-law segment in the RMS spectrum. A value of $y_r = 1$ indicates a flat $\nu F_\nu$ spectrum, where $\nu$ is the photon frequency and $F_\nu$ is the specific flux, and higher values of $y_r$ correspond to harder slopes. Specifically, for a given value of $\theta_r$, the parameter $y_r$ dictates the escape probability of photons from the RMS zone to the downstream zone. A larger value of $y_r$ means a lower escape probability and, thus, photons diffuse in the RMS zone for longer, gaining more energy on average, which results in a harder power-law segment.

The effective temperature of the hot electrons in the RMS zone, the temperature ratio between the RMS zone and the cold upstream, and the escape probability from the shock can be chosen such that the photon spectra obtained are remarkably similar to those generated by full-scale radiation hydrodynamic simulations \citep[see Figures 2 and 3 in][]{Samuelsson2022}. To account for the adiabatic compression, we use \citep{BlandfordPayneII1981}

\begin{equation}\label{eq:theta_u_K}
    \theta_{u, {\rm K}} = \left(\frac{u_u}{u_d}\right)^{1/3}\theta_u,
\end{equation}

\noindent where $u_d$ is the specific momentum of the downstream plasma in the shock rest frame. By equating the average relative energy gain per scattering, $\langle \Delta \epsilon/\epsilon \rangle$, in both models, we found in \citet{Samuelsson2022} that 

\begin{equation}\label{eq:4theta_r}
    4 \theta_r \approx \frac{(u_u)^2\ln(\bar{\epsilon}_d/\bar{\epsilon}_u)}{\xi},
\end{equation}

\noindent where $\bar{\epsilon}_u$ ($\bar{\epsilon}_d$) indicates the average photon energy in the far upstream (downstream). Here, $\xi$ is a constant and the only free parameter in the conversion. In \citet{Samuelsson2022}, we found empirically that $\xi = 55$ gave good agreement across the whole parameter space. 

Equation \eqref{eq:theta_u_K} assures that both systems get the same low-energy cutoff. Equation \eqref{eq:4theta_r} determines the high-energy cutoff of the spectrum, since the maximum energy in the shock, $\epsilon_{\rm max}$, is roughly equal to $\langle\Delta \epsilon / \epsilon\rangle$. The third and final relation is given by equating the average photon energy in the downstream, $\bar{\epsilon}_d$, between the two models. In the case of a real RMS, $\bar{\epsilon}_d$ is determined by the shock jump conditions. In \komrad, $\bar{\epsilon}_d$ is uniquely determined by the value of $y_r$ (given fixed values for $\theta_r$ and $R$). Unfortunately, there is no analytic way to relate them to one another. Hence, $\bar{\epsilon}_d$ is found in \komrad\ by numerical integration of the downstream photon distribution. 

The shape of the spectrum in the RMS zone is completely determined by the three parameters discussed above. In the downstream zone, the photons continue to interact with the electrons. The downstream electrons have a temperature $\theta_{\rm C}$, where $\theta_{\rm C}$ is the Compton temperature defined as the temperature with which there is no energy transfer between the photon and electron population. Thus, there is no energy gain in the downstream for the photon population as a whole. The Compton temperature $\theta_{\rm C}$ is not a free parameter in the model: similarly to ${\bar \epsilon}_d$, it is uniquely determined by the parameters $R$, $y_r$, and $\theta_r$.

For the downstream spectrum, the optical depth to the observer at which the shock is initiated, $\tau_i$, is an additional parameter that affects the level of thermalization that occurs in the downstream before decoupling. In \citet{Samuelsson2022}, we found that the relevant parameter determining the degree of thermalization is actually the product $\tau_i\theta_r \equiv \tau\theta$, as increasing either one increases the level of thermalization ($\tau_i$ is a measure of the number of scatterings and $\theta_r$ is a measure of the energy gain per scattering).\footnote{It would be more appropriate to say that it is the product $\tau_i\theta_{\rm C}$ that determines the level of thermalization, since $\theta_{\rm C}$ is the electron temperature in the downstream region. However, $\theta_{\rm C}$, similarly to ${\bar \epsilon}_d$, is not analytically determined by the initial model parameters and, therefore, not known before a simulation run. However, for fixed $y_r$ and $R$, the Compton temperature is directly proportional to $\theta_r$. Thus, the product $\tau_i\theta_r$ is also a measure of the level of thermalization.} Indeed, the shape of the spectrum at the photosphere is degenerate as long as $\tau\theta$, $R$, and $y_r$ are constant. Note that it is the shape of the photon spectrum that is relevant, since the bulk Lorentz factor and the luminosity of the GRB can shift the spectrum in frequency and flux. If the Lorentz factor is known, as in the case with the internal collision framework below, then the spectra are no longer degenerate and $\tau_i$ and $\theta_r$ can be decoupled.




\section{Optically shallow shocks}\label{sec:Shallow_shocks}
Typically, when RMSs are studied they are assumed to have already reached steady state \citep{LevinsonNakar2020, Samuelsson2022}. Here, we show that this is not necessarily fulfilled for a wide range of optical depths. In such cases, effects of the shock formation need to be taken into account.

The optical depth required for the radiation inside an RMS to reach steady state, $\tau_{\rm ss}$, can be approximated as the number of scatterings a cold upstream photon (after adiabatic compression) needs in order to reach the maximum energy in the RMS, $\epsilon_{\rm max}$. The average energy increase of a photon as a function of optical depth is $d\epsilon \sim \langle \Delta \epsilon \rangle \, d\tau$, where $\langle \Delta \epsilon \rangle$ is the average energy gained in each scattering. Dividing by $\epsilon$ and integrating, one gets

\begin{equation}\label{eq:tau_ss1}
    \ln \left(\frac{\epsilon_{\rm max}}{\bar{\epsilon}_{u,{\rm K}}} \right) \sim \left<\frac{\Delta \epsilon}{\epsilon}\right> \tau_{\rm ss},
\end{equation}

\noindent where $\bar{\epsilon}_{u,{\rm K}}$ is the average energy of the upstream photons after adiabatic compression. Inside an RMS, $\langle \Delta \epsilon/\epsilon\rangle$ is given by the right-hand-side of equation \eqref{eq:4theta_r}. Let us parameterize $\ln (\epsilon_{\rm max}/\bar{\epsilon}_{u,{\rm K}}) = \tilde{C} \ln (\bar{\epsilon}_d/\bar{\epsilon}_u)$. Here, $\tilde{C}$ is a proportionality constant, which for the results presented in Section \ref{sec:Results} is small, $\sim 1.5$. However, ${\tilde C}$ could potentially become large in very weak shocks were ${\bar \epsilon}_d \gtrsim {\bar \epsilon}_u$. With this parameterization, we get

\begin{equation}\label{eq:tau_ss2}
    \tau_{\rm ss} \sim \frac{\tilde{C}\xi}{(u_u)^2}.
\end{equation}

\noindent  As $\xi \approx 55$, the equation above indicates that shocks with upstream velocities $u_u \lesssim 1$ require optical depths of the order $\tau \sim 100$ to fully form. Before the shock has reached steady state, traces of the shock formation\footnote{We use the term ``shock formation'' to refer to the stages of the RMS evolution before the radiation has reached a steady state; the hydrodynamical quantities will reach steady state before this time.} may still be imprinted in the downstream spectrum. However, in this work, $u_u$ is commonly larger than unity and steady state is reached quicker, with $\tau_{\rm ss} \sim 15$.

In terms of the KRA parameters, equation \eqref{eq:tau_ss1} reads

\begin{equation}\label{eq:tau_ss3}
    \ln \left(\frac{4R}{3} \right) \sim 4\theta_r \tau_{\rm ss},
\end{equation}

\noindent where we used $\epsilon_{\rm max} \sim \langle \Delta \epsilon / \epsilon\rangle = 4\theta_r$, $\bar{\epsilon}_{u, {\rm K}} = 3\theta_{u, {\rm K}}$, as valid for a Wien distribution, and $R = \theta_r/\theta_{u,{\rm K}}$. 
In \komrad, the RMS exists for a doubling of the radius, as this is the time it takes the shock to traverse the causally connected region. This corresponds to a halving of the optical depth. An estimate of the condition required for the RMS to reach steady state is then

\begin{equation}\label{eq:tau_ss4}
    \tau\theta > \frac{\ln(R)}{2},
\end{equation}

\noindent where the $4/3$ in the logarithmic term has been dropped. 

In \citet{Samuelsson2022}, we considered shocks that occurred deep enough such that the RMS had time to reach steady state. Hence, we did not elaborate on the details regarding the shock formation. However, motivated by the fact that a wide range of optical depths can potentially result in optically shallow shocks, we have extended the KRA to more accurately capture these initial phases. The details of the extension is outlined in Appendix \ref{App:theta_r_v}. In short, we have made two changes: 1) the RMS zone now contains an initial thermal population of photons with average energy $\bar{\epsilon}_d$, and 2) the effective electron temperature in the RMS zone is allowed to vary to ensure the energy in the downstream remains constant, as imposed by the shock jump conditions. The latter results in a lower effective electron temperature initially, before it asymptotes to its steady state value $\theta_r$ as given in equation \eqref{eq:4theta_r} after around $\tau_{\rm ss}$ number of scatterings (see Figure \ref{fig:theta_r}).

\section{Methodology}\label{sec:Methodology}
The aim of this work is to find expected observational characteristics of RMSs in GRB spectra using the KRA. However, relevant ranges for the KRA parameters are not clear a priori and the resulting spectra can appear widely different depending on what initial conditions are used. 
To estimate what shock initial conditions are expected in the context of GRBs, in this work we employ a simple internal collision framework. This choice is not unique, for instance, one could instead have chosen parameter values based upon the output of numerical simulations of GRB jets. Yet, the internal collision framework is commonly used to characterize typical shock properties and often serves as a benchmark in the literature \citep[e.g.,][]{ReesMeszaros1994, Kobayashi1997, DaigneMochkovitch1998, DaigneMochkovitch2000, Lazzati1999}. It has the advantage that all relevant shock properties are analytically obtained from a few higher order parameters. Furthermore, it accounts for possible underlying correlations among the parameters, for instance, the optical depth of a collision should be correlated with the upstream temperature as they are both decreasing with increased distance to the central engine. However, the framework is highly idealised and nature is likely to be more complicated. We stress, therefore, that the intended use of the internal collision framework in this work is solely as a means to obtain relevant RMS initial conditions. 

In the next subsection, we give a brief outline of the internal collision framework (all equations used can be found in Appendix \ref{App:internal_collision}). In subsection \ref{sec:implementation}, we show how we go from the higher order internal collision parameters to the KRA parameters. A KRA spectrum is generated from the parameters, forward-folded through the \textit{Fermi} response, and fitted with a CPL function as explained in subsection \ref{sec:synthetic_spectra}.

\subsection{Subphotospheric internal collision framework}\label{sec:internal_collision}
In this work, we use a simple internal collision framework to estimate the relevant parameter space for RMSs in GRBs. A detailed description of the internal collision framework used is given in Appendix \ref{App:internal_collision}. Most of the equations in Appendix \ref{App:internal_collision} can be found in relevant reviews on GRBs such as \citet{Meszaros2006, Peer2015}. In this subsection, we give a brief overview.

We consider a variable central engine that launches a slower and a faster fireball with terminal Lorentz factors $\Gamma_s$ and $\Gamma_f \equiv \psi\Gamma_s$, respectively, where $\psi > 1$ is a free parameter. The two fireballs are launched with a time separation $\delta t$, which we parametrize as $\delta t = \chi r_0/c$. Here, $\chi$ is a constant and $r_0$ is the radius where the free fireball acceleration begins. This parametrization implies that we assume the time separation to be proportional to the light crossing time as the base of the jet. Throughout this work, we use $\chi = 1$, however, in principle $\chi$ could be both smaller or larger \citep{ReesMeszaros2005}. The initial Lorentz factor of the fireballs at the base of the jet, $\Gamma_0$, is also a free parameter, although its value has little effect on the final result.

The speed difference between the two fireballs eventually leads to an internal collision at some radius $r_{\rm coll}$. Given that $\psi \gg 1$, the collision radius is given by $r_{\rm coll} = 2\Gamma_s^2\chi r_0$. The collision will be subphotospheric as long as $r_{\rm coll}$ is smaller than the photospheric radius of the slower fireball, $r_{{\rm ph},s}$. As we are only interested in subphotospheric collisions, we use the optical depth for the slower fireball at which the collision occurs, $\tau_s$,
as a free parameter. For a conical outflow, one has $\tau_s = r_{{\rm ph},s}/r_{\rm coll}$. When the collision occurs, a reverse and a forward RMS are launched into the two fireballs and the shocked plasma is accumulated in a common downstream in between the two fireballs. Somewhat surprising is that the photospheric radius for the common downstream is slightly larger compared to that of the slower fireball due to shock compression of the plasma, even though the downstream has an increased velocity (see equation \eqref{eq:tau}). 

When the RMS gets closer to the photosphere, photons start to leak out of the shock and the decreasing radiation pressure leads to the formation of two collisionless shocks \citep{Peer2008, Beloborodov2011, LundmanBeloborodov2021}. As \komrad\ cannot model these higher order effects, we restrict ourselves to shocks where $\tau_s > 5$ \citep[see][]{Beloborodov2011}. With this choice, we find that the corresponding optical depth for the downstream, $\tau_i$, is always larger than 10. After the collision, the energized radiation is contained in the downstream, which travels at a bulk Lorentz factor $\Gamma = \sqrt{\Gamma_s\Gamma_f} = \Gamma_s\sqrt{\psi}$.

\begin{figure}
    \centering
    \includegraphics[width=\columnwidth]{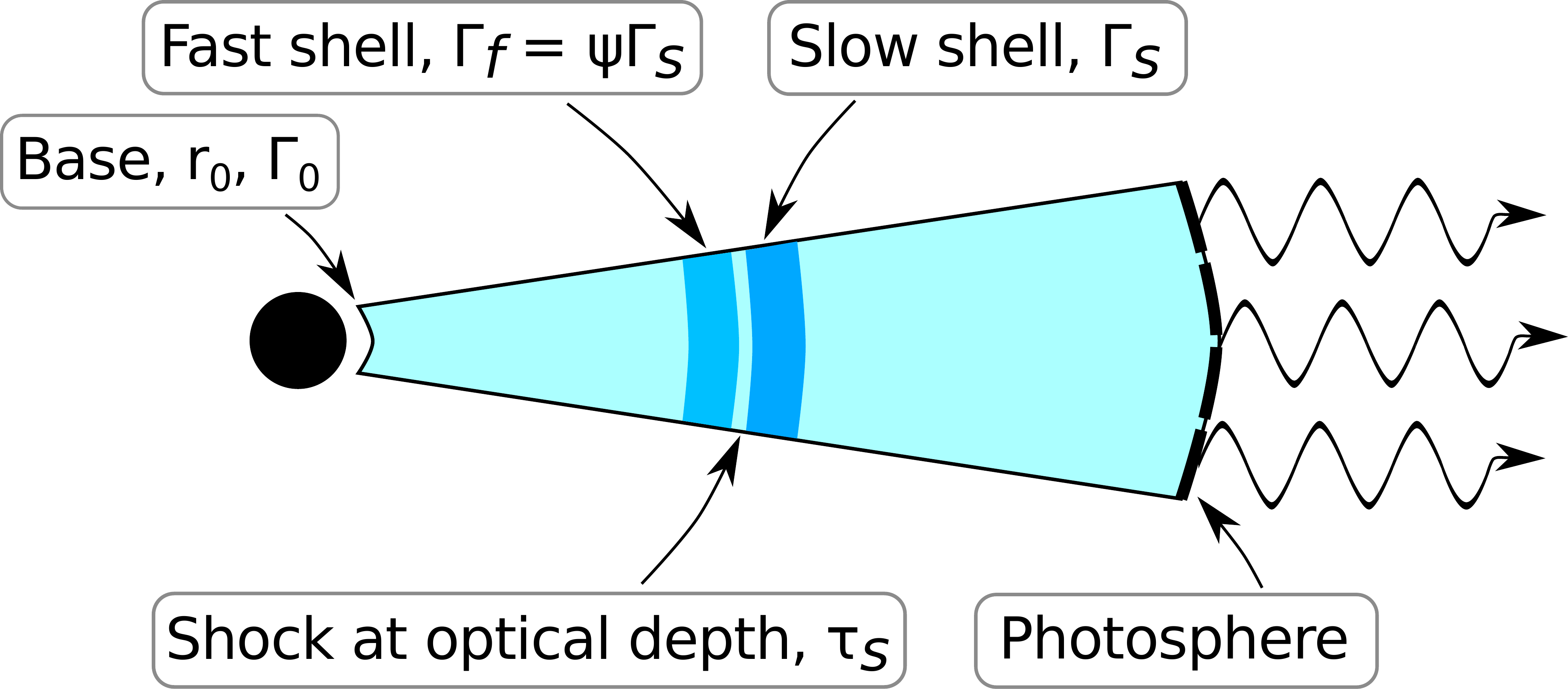}
    \caption{A schematic of the subphotospheric internal collision framework. Two fireballs are accelerated from the base in a conical outflow, starting at radius $r_0$ and with an initial Lorentz factor $\Gamma_0$. The slower of the two fireballs has a bulk Lorentz factor $\Gamma_s$ and the faster fireball has a Lorentz factor $\Gamma_f = \psi \Gamma_s$. The collision occurs at an optical depth (as measured for the slower fireball) $\tau_s$, with the requirement $\tau_s>5$ to avoid the dynamical changes that occur in the RMS at small optical depths.}
    \label{fig:collision_schematic}
\end{figure}

A schematic of the internal shock framework is shown in Figure \ref{fig:collision_schematic}. To minimize the number of free parameter, we fix the values of $r_0$ and $\Gamma_0$ and study only two different jet base configurations (see Section \ref{sec:Results} for motivation and further discussion). We title these two scenarios the \textit{early acceleration scenario} with $r_0 = 10^7~${cm} and $\Gamma_0 = 1.5$ and the \textit{delayed acceleration scenario} with $r_0 = 10^{10}~${cm} and $\Gamma_0 = 4$ .

\subsection{From the internal collision parameters to the KRA parameters}\label{sec:implementation}

To arrive at the results presented in Section \ref{sec:Results}, we start by randomly assigning values to $\log(\Gamma_s)$, $\log(\tau_s)$, and $\psi$ using normal distributions as given in Table \ref{tab:init_dist}.
Each parameter has a minimum requirement as $\Gamma_s > 5$, $\tau_s > 5$, and $\psi > 2$, which collectively removes $\sim 1$\% of generated parameter triplets. For each valid triplet, we calculate $\theta_{u}$ and $\tilde{n}$ using equations \eqref{eq:theta_u2} and \eqref{eq:tilde_n}.\footnote{The values for $\theta_u$ and ${\tilde n}$ used for the results presented in Section \ref{sec:Results} are those obtained for the slower fireball, which are slightly different compared to the values obtained for the faster fireball as discussed in Appendix \ref{App:internal_collision}. Using the values from the faster fireballs results in a softer spectrum with a lower peak energy, see Figure \ref{fig:different_params_reverse_shock}.} With the velocity transformations $\gamma_u = (1-\beta_s\beta_{\rm fs})\Gamma_s\Gamma_{\rm fs}$ and $\gamma_d = (1-\beta\beta_{\rm fs})\Gamma\Gamma_{\rm fs}$, where $\beta_{\rm fs}$ is the dimensionless velocity of the forward shock as measured in the lab frame and $\Gamma_{\rm fs} = (1-\beta_{\rm fs}^2)^{-1/2}$, we find $u_u$, $u_d$, and $\bar{\epsilon}_d$ using the shock jump conditions \citep[e.g.,][]{Beloborodov2017, Samuelsson2022}. We only include runs where $u_u < 3$ in the final results, as above this limit, relativistic effects such as pair production, Klein-Nishina cross section, and anisotropy start to have an effect \citep{Samuelsson2022}. This removes $\lesssim 4$\% of remaining spectra. 



With $u_u$, $u_d$, and $\bar{\epsilon}_d$ determined, one can obtain the KRA parameters $\tau\theta \equiv \tau_i \theta_r$ and $R$ using equations \eqref{eq:theta_u_K}, \eqref{eq:4theta_r}, and $\eqref{eq:tau}$. In the two table models that are used to generate the synthetic spectra (one for the early and one for the delayed acceleration scenario, see next subsection), we set bounds as $1 < \tau\theta < 50$ to increase the resolution of $\tau\theta$ around the most relevant parameter region. Less than 3\% of remaining spectra have values of $\tau\theta$ outside of these limits, and are thus removed. Bounds are similarly set for $R$ and $y_r$ in the table model, which are different in the the early and delayed acceleration scenarios, but these have a minor effect. All in all, $\sim 93$\% of initially generated parameter triplets survive all cuts.

Unfortunately, there is no exact analytical expression that relates $y_r$ to $\bar{\epsilon}_d$. Preferably, one would iteratively run \komrad\ (without adiabatic cooling as this is not accounted for in the jump conditions) and integrate the downstream spectrum until one finds the unique value of $y_r$ that, for the current value of $R$ and $\theta_r$, gives an average downstream energy of $\bar{\epsilon}_d$ (note that ${\bar \epsilon}_d$ is independent of $\tau_i$ when there is no adiabatic cooling). However, this is computationally too costly. Instead, we create a grid with 27\,000 spectra (30 values each for $\theta_r$, $R$, and $y_r$, neglecting adiabatic cooling) over the relevant parameter space and save $\bar{\epsilon}_d$. We then find the spectrum in the grid that has the closest values to the current $\theta_r$, $R$, and $\bar{\epsilon}_d$, and use the value of $y_r$ from that simulation as the true value. We use different grids for the early and the delayed acceleration scenarios to maximize resolution. This prescription results in a maximum error in $y_r$ of 0.09 and 0.05 for the early and delayed acceleration scenario, respectively. We ensure that a recovered value of $y_r$ is never on a grid border. 

\begin{table}[]
    \centering
    \caption{Mean and standard deviations for the normal distributions used to generate values for the internal collision parameters. The logarithms used for $\Gamma_s$ and $\tau_s$ are base 10.}
    \begin{tabular}{ccc}
    \hline
         Parameter & Mean & Standard deviation  \\
    \hline
         $\log(\Gamma_s)$ & 1.5 & 0.2 \\
         $\log(\tau_s)$ & 1.5 & 0.3 \\
         $\psi$ & 15 & 5 \\
    \hline
    \end{tabular}
    \label{tab:init_dist}
\end{table}

\subsection{Synthetic spectrum and fit}\label{sec:synthetic_spectra}
Having obtained the KRA parameters, we generate a synthetic \komrad\ spectrum using the Multi-Mission Maximum Likelihood Framework \citep[3ML]{Vianello2015}. In 3ML, we have constructed two table models consisting of 1\,000 \komrad\ spectra in the early acceleration scenario (10 values each for $\tau\theta$, $R$, and $y_r$) and 5\,586 spectra in the delayed acceleration scenario (21, 14, and 19 values for $\tau\theta$, $R$, $y_r$, respectively). The two table models have different ranges of $R$ and $y_r$ to maximize resolution. With the given KRA parameter values, 3ML interpolates between the values in the table model to generate the synthetic spectrum. All spectra in the table models have been broadened to account for high latitude emission and different radial contributions to the photospheric emission, as discussed in Appendix \ref{App:broadening}. The broadened spectrum is Doppler boosted into the observer frame using the correct Doppler factor obtained in Appendix \ref{App:broadening}. Also accounting for redshift, each photon is blue-shifted by a factor $5\Gamma/3(1+z)$. We use $z = 1$ throughout this work.

We account for background and detector effects by forward folding the model spectrum through the response of \textit{Fermi}, using a response file from an arbitrary burst, here taken to be GRB 150213A. We verified that the specific response does not have an impact on the results by trying several other ones from different GRBs. This means that we are also considering background in our analysis. However, to more clearly observe the features of the signal, we require the synthetic data to have a signal-to-noise ratio (SNR) above 30. The SNR for each generated spectrum varies, and the median is $\sim 45$. 

Once the synthetic spectrum is obtained, we perform a standard \textit{Fermi} data analysis \citep[e.g.,][]{Goldstein2012, Yu2019}. In particular, we fit the synthetic data with a CPL function using the Maximum Likelihood framework in 3ML. This analysis yields values for $\alpha$ and $E_{\rm peak}$, where $E_{\rm peak} = (\alpha+2)E_c$ is the peak energy of the CPL function in a $\nu F_\nu$ spectrum.


\section{Results}\label{sec:Results}
As mentioned in Section \ref{sec:Methodology}, in this work we use two different jet base configurations, characterized by their different values of $r_0$ and $\Gamma_0$. In the first, the free acceleration starts close to the central engine with $r_0 = 10^7~$cm and $\Gamma_0 = 1.5$. Similar conditions have been invoked or found through data analysis in earlier studies and are therefore interesting to investigate \citep{ReesMeszaros1994, MeszarosRees2000, Iyyani2013, Larsson2015}. However, hydrodynamical simulations show that the jets in both long and short GRBs typically remain collimated due to the high pressure of the surrounding material \citep{Lazzati2009, Lopez-Camara2013, Gottlieb2018, Gottlieb2019, Gottlieb2021}. The collimation leads to shocks that suppress the acceleration, effectively moving the base of the conical jet much further out. Motivated by these results, we use $r_0 = 10^{10}~$cm and $\Gamma_0 = 4$ for our second scenario \citep[from a long GRB simulation in][]{Gottlieb2019}. That the acceleration starts around $r_0 = 10^{10}~$cm was also suggested theoretically by \citet{Thompson2007}. We call the two scenarios the \textit{early} and the \textit{delayed} acceleration scenario, respectively. In both cases, we use $\chi = 1$, however, in principle $\chi$ could be both larger or smaller than unity depending on the dynamics at the base of the jet \citep{ReesMeszaros2005}.

\begin{figure}
    \centering
    \includegraphics[width=\columnwidth]{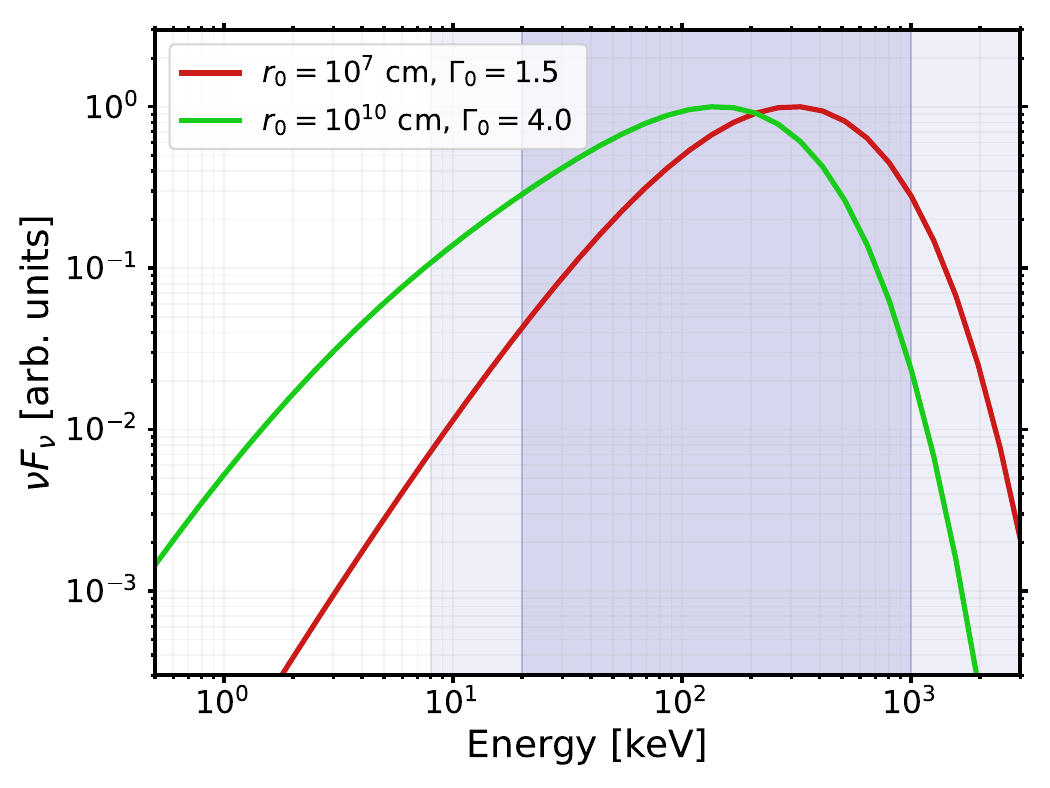}
    \caption{Photospheric KRA spectra generated using \komrad\ in the internal collision framework for $\Gamma_s = 10^{1.5}$, $\tau_s = 10^{1.5}$, and $\psi = 10$ in the early acceleration scenario (red, $r_0 = 10^7~$cm, $\Gamma_0 = 1.5$) and in the delayed acceleration scenario (green, $r_0 = 10^{10}~$cm, $\Gamma_0 = 4$). The spectra are shown prior to the forward-folding through the \textit{Fermi} response. The shaded area is a rough indication of the \textit{Fermi} GBM sensitivity range, with lighter shading implying less sensitivity \citep{Meegan2009}. The parameter values obtained in the early acceleration scenario are $\tau\theta = 4.2$ ($\tau = 95$, $\theta_r =0.045$), $R = 27$, and $y_r = 3.76$, and $\Gamma = 100$. In the delayed acceleration scenario, they are $\tau\theta = 5.1$ ($\tau = 97$, $\theta_r =0.053$), $R = 220$, $y_r = 2.1$, and $\Gamma = 100$.}
    \label{fig:typical_spectra}
\end{figure}


\begin{figure}
    \centering
    \includegraphics[width=\columnwidth]{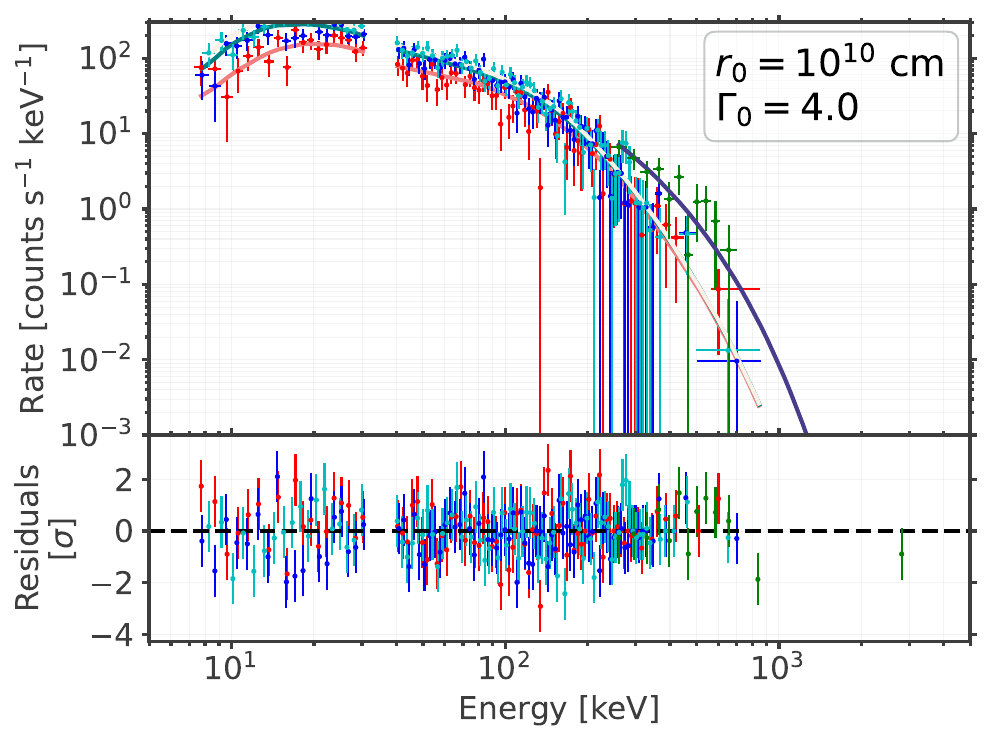}\\
    \includegraphics[width=\columnwidth]{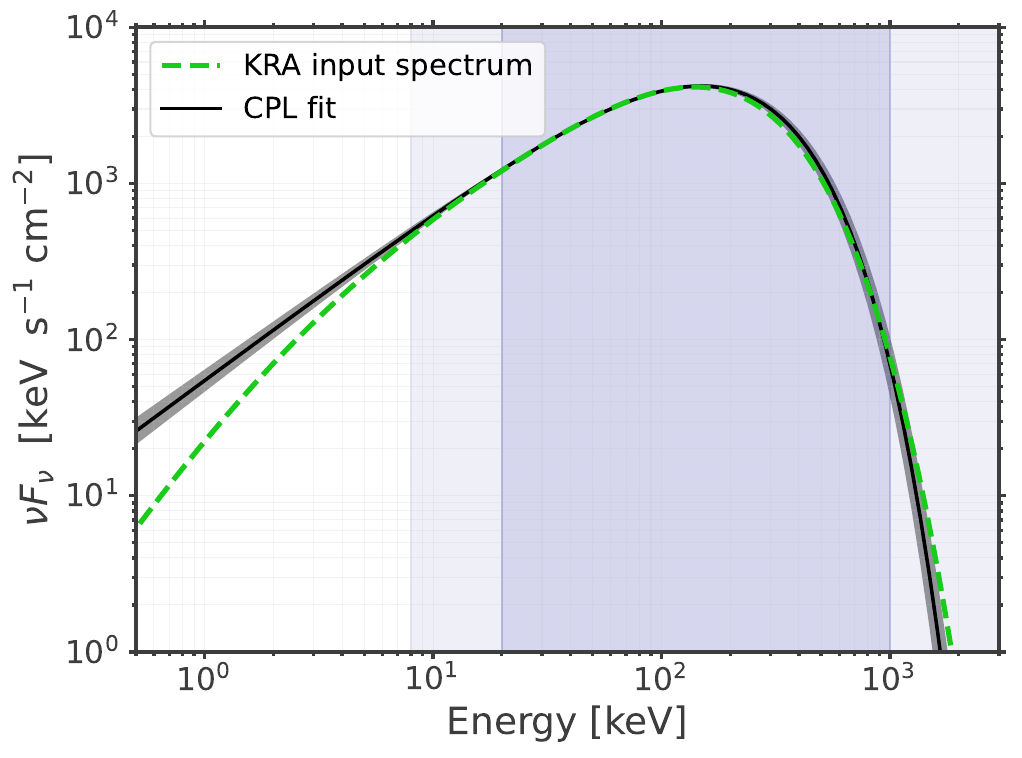}
    \caption{Count spectrum (top) and $\nu F_\nu$ spectrum (bottom) obtained by fitting a CPL function to the (forward-folded) photospheric spectrum in the delayed acceleration scenario from Figure \ref{fig:typical_spectra}. As the hardening of the input spectrum at low energies is outside of the GBM sensitivity range (indicated by the shaded area), the fit is adequate. In the bottom panel, the un-folded photospheric spectrum from Figure \ref{fig:typical_spectra} has been overlaid for comparison. 
    Best fit parameters are $\alpha = -0.92 \pm 0.05$ and $E_{\rm peak} = 150_{-12}^{+13}~$keV.}
    \label{fig:example_fit}
\end{figure}

\subsection{General spectral characteristics}\label{sec:spectral_characteristics}

In Figure \ref{fig:typical_spectra}, we show the photospheric spectra produced by \komrad\ in the early acceleration scenario in red and the delayed acceleration scenario in green for $\Gamma_s = 10^{1.5}$, $\tau_s = 10^{1.5}$, and $\psi = 10$. In the early acceleration scenario, the spectrum is hard. This is partly because the photon-to-proton ratio, $\tilde{n}$, increases with $r_0$ as evident from equation \eqref{eq:tilde_n}. Thus, for small values of $r_0$, fewer photons share the same dissipated energy, leading to a steeper power-law slope (larger values of $y_r$). The hardness also increases due to the ratio, $R$, between the upper and lower cutoff shrinking with decreasing $r_0$, as indicated by equation \eqref{eq:R}. A shorter power-law segment leads to a quicker thermalization. That fewer photons share the available energy also pushes $E_{\rm peak}$ to larger values, as the average photon energy in the comoving frame is higher.

In the delayed acceleration scenario, the slope of the spectrum is softer. The smooth curvature exhibits a double power-law behavior at low energies. The main power-law component ($\sim 10{\rm -}100~$keV) is the remainder of the power-law segment generated in the RMS. The low-energy power law ($\lesssim 3~$keV) is the Rayleigh-Jeans slope from the initial thermal population, softened due to high-latitude emission and radial contributions to the photosphere, as explained in Appendix \ref{App:broadening}. In the spectrum in Figure \ref{fig:typical_spectra}, the low-energy power law is below the energy range of GBM. 

In \citet{Samuelsson2022}, we argued that the observed peak energy could correspond either to the low-energy cutoff or to the high-energy cutoff in the spectrum, as we were agnostic to the value of $y_r$. However, due to the rather small values of $\tilde{n}$ obtained when employing the internal collision framework, we find that $y_r>1$ is essentially always satisfied. This implies a positive slope in a $\nu F_\nu$ spectrum, similar to those shown in Figure \ref{fig:typical_spectra}. Thus, the peak energy corresponds to the high-energy cutoff, which is the maximum energy in the spectrum after thermalization and adiabatic cooling. Therefore, in the current model, we always expect a sharp drop in the spectrum above the observed peak. To obtain a high-energy power law, additional mechanisms must be considered such as particle acceleration in subshocks in magnetic or relativistic RMSs \citep{LundmanBeloborodov2019, Levinson2020}, or via shear interactions between the jet and the surrounding matter \citep{Ito2013, VyasPeer2022}.

The additional break at lower energies in the delayed acceleration spectrum corresponds to the initial cold upstream photons, after having been subjected to thermalization and adiabatic cooling. A lower limit for the additional break is $\theta_{u, {\rm K}}^{\rm ph}$, defined as the temperature a cold photon would have at the photosphere if one ignored any heating via scatterings. Accounting for adiabatic cooling to the photosphere, one gets $\theta_{u,{\rm K}}^{\rm ph} = \theta_{u,{\rm K}}\times (6/5)\tau_i^{-2/3}$, where $\tau_i^{-2/3}$ accounts for the idealised adiabatic cooling in the optically thick regime and the coefficient $(6/5)$ appears as to not overestimate the cooling \citep[][see Appendix \ref{App:broadening}]{Peer2008,Beloborodov2011}. This translates into an expected additional observed break in the spectrum at a temperature above

\begin{equation}
\begin{split}
    k T_{\rm break} &> \frac{5\Gamma}{3(1+z)} \theta_{u, {\rm K}}^{\rm ph} m_e c^2 \\[2.2mm] 
    \approx 0.7~{\rm keV} &~ \chi^{-5/12} \, r_{0,10}^{-1/4} \, \beta_0^{-1/4} \, \Gamma_0^{-1/6} \, \Gamma_{s,1.5}^{7/12}\, \tau_{s,1.5}^{-5/12} \, \psi_1^{2/3}.
\end{split}
\end{equation}

\noindent where we used equations \eqref{eq:theta_u2}, \eqref{eq:Gamma}, \eqref{eq:tau}, \eqref{eq:theta_u_K2}. For the spectrum in the delayed acceleration scenario shown in Figure \ref{fig:typical_spectra}, the break occurs at $\sim 5~$keV. 
An additional break in the low-energy part of the spectrum is interesting as it has been observed in several studies \citep[][see also subsection \ref{sec:disc_2SBPL}]{Stroh1998, Ravasio2019, Burgess2020, Gompertz2023}.


In Figure \ref{fig:example_fit}, we show the delayed acceleration scenario spectrum from Figure \ref{fig:typical_spectra} fitted with a CPL function. The KRA spectrum has been forward-folded through the \textit{Fermi} response to account for background and detector effects. The top panel shows the count spectrum, while the bottom panel shows the $\nu F_\nu$ spectrum. In this case, the hardening at low energies is below the sensitivity window of GBM. Thus, the CPL function provides an adequate fit to the data, as can be seen from the residuals. 



\begin{table*}
    \centering
    \caption{Mean and standard deviations for the RMS and the KRA parameters used to obtain the CPL parameter distributions in Figure \ref{fig:hist_CPL}. The RMS and KRA parameters in turn are obtained using the internal collision framework as described in Section \ref{sec:Methodology} and Appendix \ref{App:internal_collision}, with the initial parameter distributions given in Table \ref{tab:init_dist}.}
    \begin{tabular}{cccccccc}
    \hline
          & $\theta_{u} ~ [10^{-4}]$& ${\tilde n} ~ [10^5]$ & $u_u$ & $\tau_i$ & $\theta_r ~ [10^{-2}]$ & $R$ & $y_r$ \\
    \hline
        Early acceleration 
        & $8.07\pm 2.0$
        & $0.28\pm 0.053$ 
        & $2.24\pm 0.41$
        & $121\pm 94$ 
        & $8.30 \pm 3.3$ 
        & $49.2\pm 23$ 
        & $3.81\pm 0.82$ \\
        Delayed acceleration
        & $1.15\pm 0.26$
        & $1.03\pm 0.23$ 
        & $2.19\pm 0.46$ 
        & $120\pm 75$ 
        & $9.41 \pm 4.2$ 
        & $389\pm 197$ 
        & $2.24\pm 0.22$ \\
    \hline
    \end{tabular}
    \label{tab:parameter_dist}
\end{table*}

\begin{figure*}
    \centering
    \includegraphics[width=\columnwidth]{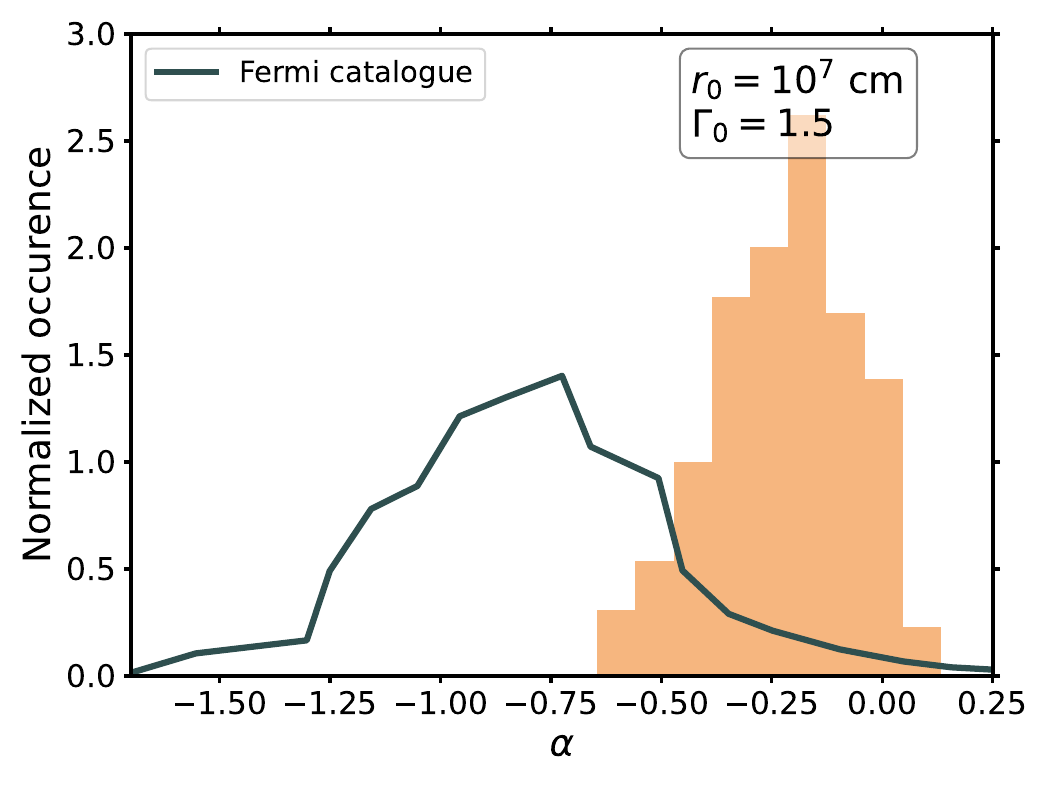}
    \includegraphics[width=\columnwidth]{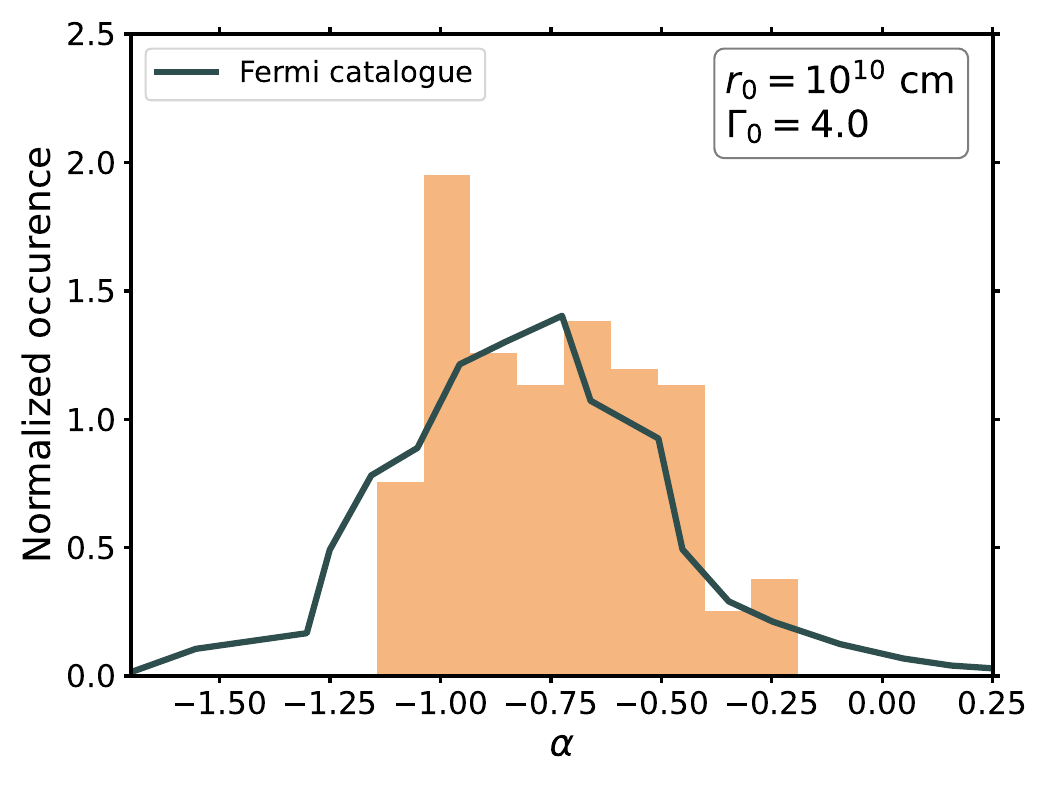}\\
    \includegraphics[width=\columnwidth]{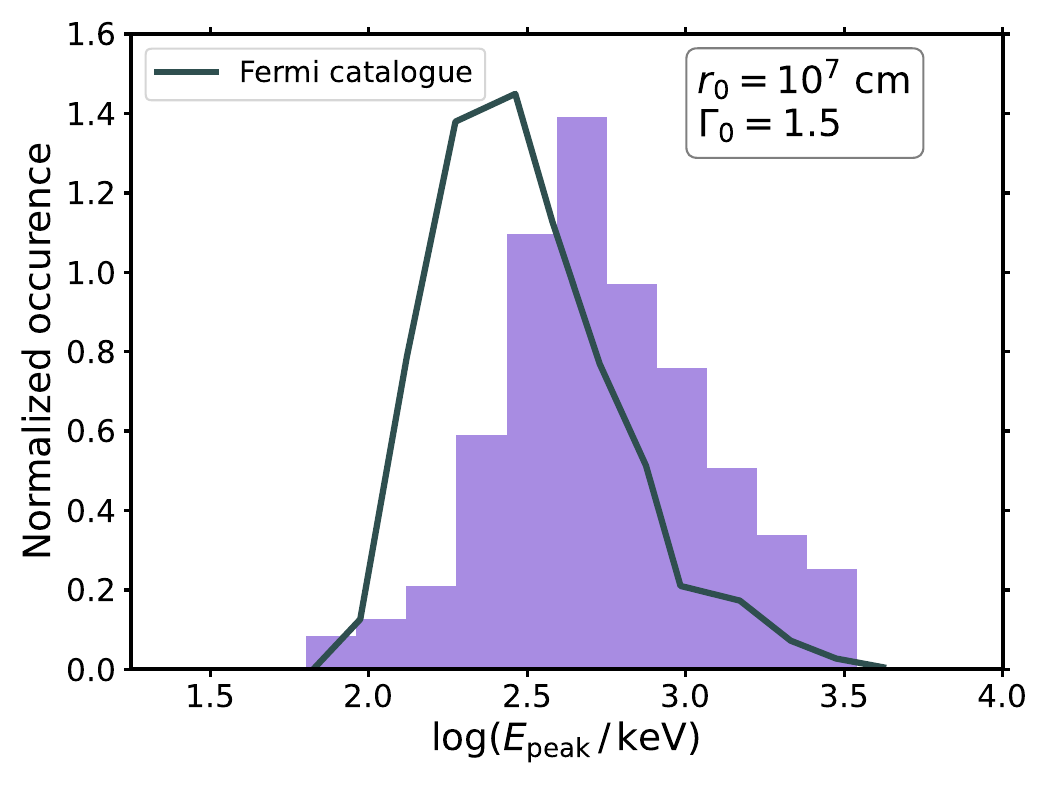}
    \includegraphics[width=\columnwidth]{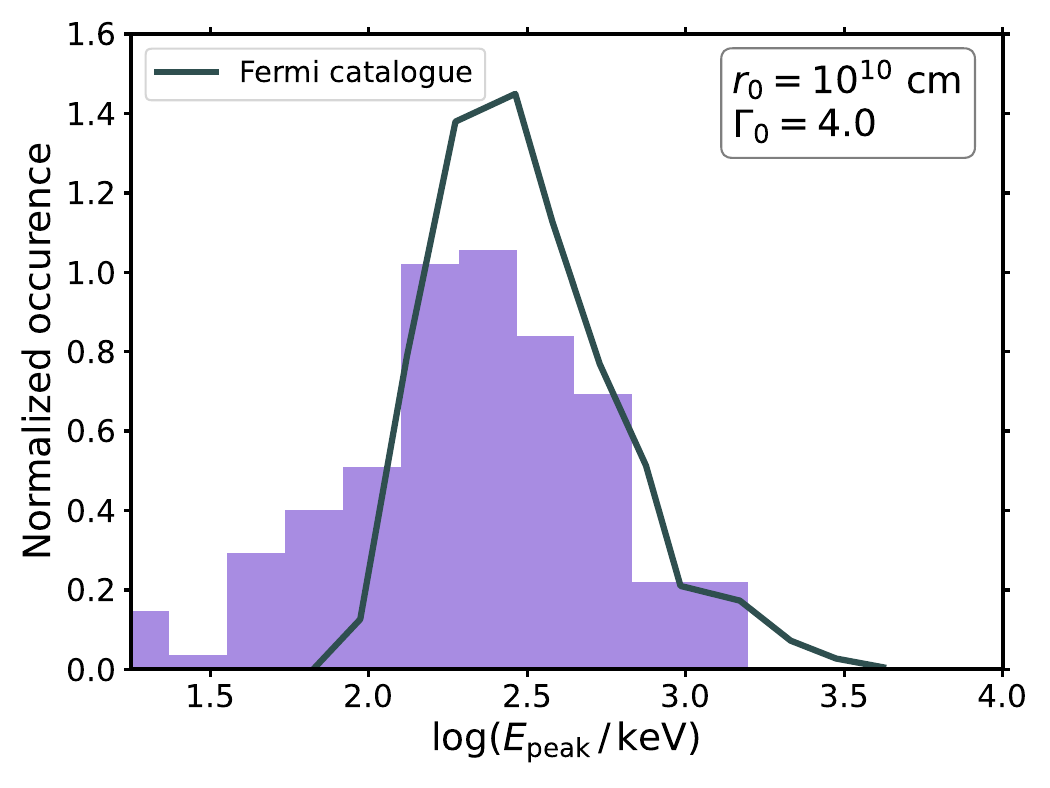}
    \caption{Histograms obtained by fitting a CPL function to 150 synthetic photospheric spectra, using the method described in subsections \ref{sec:implementation} and \ref{sec:synthetic_spectra}. The mean and standard deviations for the input RMS and KRA parameters are given in Table \ref{tab:parameter_dist}, which in turn are obtained from the higher order initial parameter distributions given in Table \ref{tab:init_dist}. Top panels show $\alpha$ and bottom panels show $E_{\rm peak}$, for the early acceleration scenario (left) and the delayed acceleration scenario (right), as also indicated in the panels. The solid dark green lines show the distributions found for the CPL parameters in a time-resolved catalogue of bright \textit{Fermi} GRBs during the first four years of observations \citep{Yu2016}. It is evident that unless collimation shocks delay the acceleration, the distribution obtained is harder than most observed bursts. 
    All panels are made with $\chi = 1$ and $z = 1$.}
    \label{fig:hist_CPL}
\end{figure*}

\subsection{Parameter distributions}\label{sec:parameter_dist}
In Figure \ref{fig:hist_CPL}, we show histograms of parameter values obtained from 150 CPL function fits to synthetic KRA spectra, using the method described in subsections \ref{sec:implementation} and \ref{sec:synthetic_spectra}. The input distributions for $\Gamma_s$, $\tau_s$, and $\psi$ are given in Table \ref{tab:init_dist} (in Appendix \ref{App:different_params}, we show results for two additional input distributions). The corresponding mean values and standard deviations for the RMS and the KRA parameters are given in Table \ref{tab:parameter_dist}. The generated histograms of $\alpha$ (top) and $E_{\rm peak}$ (bottom) are compared to the corresponding distributions found in a time-resolve analysis of bright \textit{Fermi} bursts during the first four years of observations \citep{Yu2016}.

Generally, the fits obtained by the CPL function are good, with residuals similar to those shown in the top panel of Figure \ref{fig:example_fit}. However, the CPL function cannot quite capture the smooth curvature of the photospheric spectra around the peak. 
This leads to the CPL function sometimes underestimating the flux at the highest energies, but this depends on the position of the peak energy and the SNR of the input spectrum. 
In these cases, the Band function may be a better fit. Furthermore, the CPL function cannot account for the hardening of the spectrum that occurs at lower energies in the delayed acceleration scenario. For high values of $\Gamma$, this hardening is within the GBM energy range. In these cases, adding an additional power law to the fitting function significantly improves the fit (see subsection \ref{sec:disc_2SBPL}).

\subsubsection{Early acceleration scenario}
The obtained parameter distributions in the early acceleration scenario are shown on the left-hand-side in Figure \ref{fig:hist_CPL}. It is evident from the figure that in the absence of collimating material that delays the acceleration, the generated spectra are too hard to account for the majority of observed bursts. 
Furthermore, the $E_{\rm peak}$ distribution is a factor of a few too high compared to the observations. Both of these findings are in agreement with the discussion in subsection \ref{sec:spectral_characteristics} and with the single spectrum shown for the early acceleration scenario in Figure \ref{fig:typical_spectra}. It can also be deduced from Table \ref{tab:parameter_dist}, which shows that both ${\tilde n}$ and $R$ are significantly smaller compared to the delayed acceleration scenario. 
However, we note that the value of $E_{\rm peak}$ is dependent on the redshift, which is here fixed to $z=1$. The power-law slopes in the early acceleration scenario is consistent with roughly a quarter of the time resolved fits in \citet{Yu2016}, which have $\alpha \gtrsim -0.6$.

The early acceleration parameter distributions in Figure \ref{fig:hist_CPL} are similar to those found for the non-dissipative photospheric model studied in \citet{Acuner2019}. The reason is that the spectral slope is largely determined by the broadening effects discussed in Appendix \ref{App:broadening} and any observable trace of the dissipation is mostly absent. Thus, GRB spectra that are well fitted with a non-dissipative photospheric model are not necessarily without prior dissipation. Dissipation could have been present, but its effect on the final spectrum may be obscured by the geometrical broadening effects.

\subsubsection{Delayed acceleration scenario}
The results in the delayed acceleration scenario are shown on the right-hand-side in Figure \ref{fig:hist_CPL}. When a collimation shock delays the acceleration, the range of $\alpha$-values obtained is more similar to the observed range, with typical values centered around $\alpha = -0.8$. This shows that dissipative photospheric emission, when accounting for the correct shock physics below the photosphere, can be much softer than what is commonly assumed for photospheric spectra in the literature. 

The obtained distribution for the peak energy has a median value of $\sim 220$ keV. However, it overestimates the number of bursts with low values of $E_{\rm peak}\lesssim 100$. This is because we artificially increase the SNR to be above 30 in our modelling. In reality, there is an observational bias against bursts with lower values of $E_{\rm peak}$, which is unaccounted for here. Furthermore, we again note that the $E_{\rm peak}$ distribution is sensitive to the value of the redshift used.

\section{Underlying assumptions}\label{sec:Assumptions}

\subsection{Different initial distributions}\label{sec:different_distributions}
The shape of the histograms presented in Figure \ref{fig:hist_CPL} depend on the choice of the initial parameter distributions shown in Table \ref{tab:init_dist}. To check the robustness of our results, we present similar figures for two different input parameter distributions in Appendix \ref{App:different_params}. Furthermore, we show the effect of using
the values of ${\tilde n}$ and $\theta_u$ obtained for the faster fireball instead of the slower fireball in equations \eqref{eq:theta_u2} and \eqref{eq:tilde_n}.

It is evident from the figures that the detailed appearance of the histograms changes in these cases. This highlights the importance of the chosen methodology on the final results. However, the conclusions that dissipative photospheric emission can be very soft and closely resemble the spectral shapes that are observed remain robust. Indeed, when using the conditions in the faster instead of the slower fireball, values as low as $\alpha < -1.5$ are obtained in a few cases. This is mainly due to the increase in the photon-to-proton ratio for the same initial internal collision parameters.

\subsection{Additional emission channels}\label{sec:emission_channels}
In this work, we have studied internal collisions that result in shocks in the optically thick regions. However, the optical depth of an internal collision is set by the properties of the central engine, and there is no a priori reason to believe all collisions would be subphotospheric. Including shocks above the photosphere would alter the $\alpha$-histograms shown in Figure \ref{fig:hist_CPL}. Specifically, synchrotron emission from fast cooling electrons results in softer values of $\alpha \sim -3/2$ \citep{Sari1998, Burgess2015}. Additionally, photospheric emission without dissipation, or with dissipation but small values of $r_0$, give hard values of $\alpha > -0.5$. Moreover, under certain conditions, such as if the jet becomes transparent during the acceleration phase, values up to $\alpha \sim 0.7$ can be reached \citep{Ryde2017, Parsotan2018, Acuner2019}. Inspection of the delayed acceleration scenario in Figure \ref{fig:hist_CPL} shows that we currently underestimate the frequency of bursts with $\alpha$-values in these regions. Therefore, it may be that all three channels combined make up the observed histogram. 

A model that can properly account for collisions both above and below the photosphere would be an interesting future study. However, we note that the apparent smoothness of the $\alpha$-distribution may favor a single emission mechanism. 
It is possible that for a suitable superposition of the histograms presented in Figure \ref{fig:hist_CPL} together with the ones presented in Appendix \ref{App:different_params}, the current framework could account for the entire $\alpha$-range. Moreover, allowing $r_0$ to vary between bursts would also broaden the $\alpha$-distribution. A more realistic model can also include different observation angles, multiple internal collisions, and/or magnetic fields, as well as allowing for data with lower significance. However, these claims require future study.


\subsection{Validity of KRA}\label{sec:assumptions_KRA}
The KRA assumes negligible magnetic fields in the region where the RMS occurs. As magnetic fields likely play a crucial role in the launching of the jet, this assumption depends on how quickly the initial magnetic field is dissipated. If there exists moderate magnetization across the RMS transition layer, a collisionless sub-shock can form in the immediate downstream \citep{LundmanBeloborodov2019}. Charged particles can be accelerated in the sub-shock leading to enhanced photon production via synchrotron emission. This can shift the photon-to-proton ratio $\tilde{n}$ to higher values, possibly increasing the viable parameter space of smaller values of $r_0$ in our model. 

In addition, the KRA cannot account for pair production or shock breakout emission. Violent pair production can occur in the downstream of an RMS if the initial relativistic velocity of the upstream is large enough \citep[$u_u \gtrsim 3$,][]{Lundman2018}. These conditions may be difficult to obtain in an internal shock framework as it requires large differences between the terminal outflow velocities in different parts of the jet. If present, pair creation can greatly alter the spectrum generated in the downstream \citep{Katz2010, Budnik2010, Beloborodov2013, Lundman2018}. 

Shock breakout in the context of GRBs is an interesting topic that has seen a lot of recent development, specifically in the context of GRB 170817A \citep[e.g.,][]{Bromberg2018, Gottlieb2018, LundmanBeloborodov2021}.
When the shock gets close to the photosphere, it broadens and photons start leaking out of the RMS which decreases the radiation pressure until two collisionless subshocks form \citep{Levinson2012, LundmanBeloborodov2021}. Therefore, the dynamics are more complicated, which will affect the resulting spectrum.

\section{Discussion}\label{sec:Discussion}


\begin{figure*}
    \centering
    \includegraphics[width=\columnwidth]{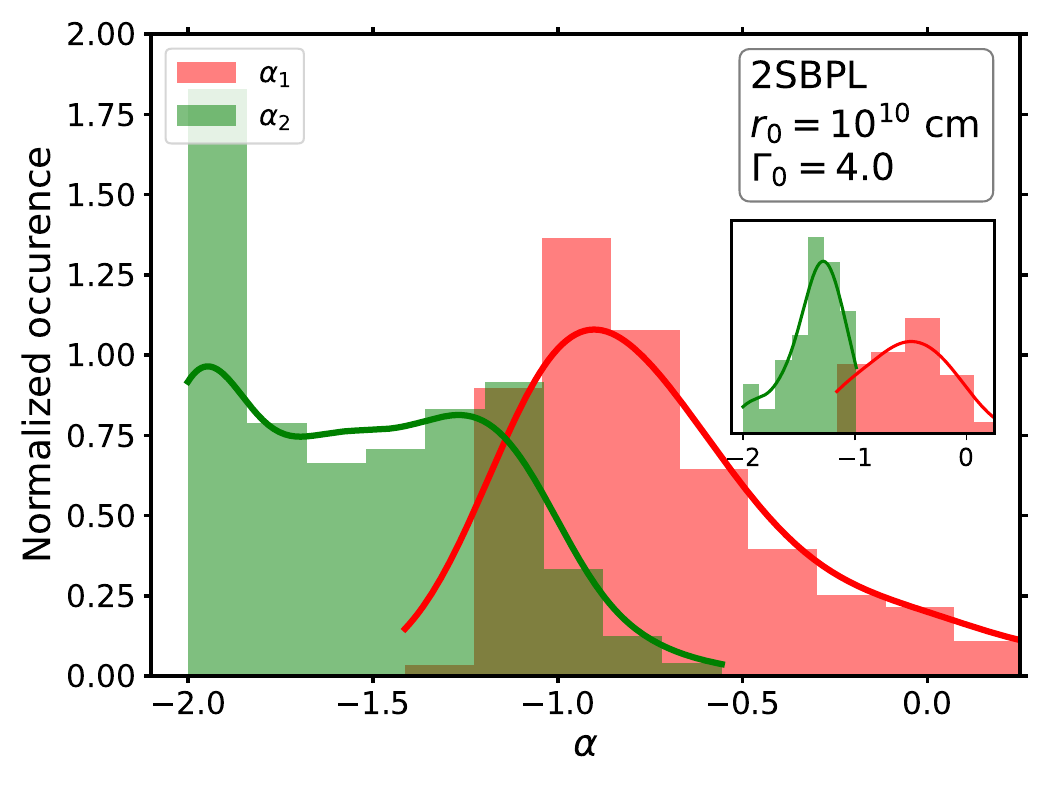}
    \includegraphics[width=\columnwidth]{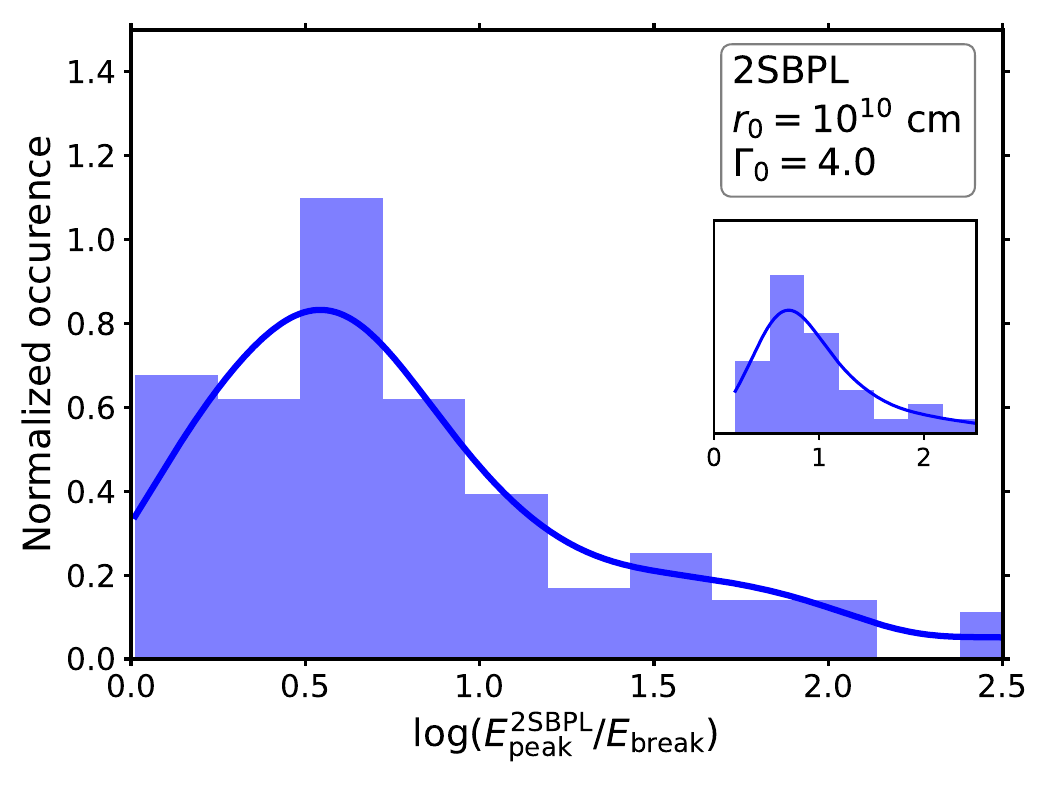}
    \caption{Histograms obtained by fitting a 2SBPL function to 150 synthetic photospheric spectra in the delayed acceleration scenario, obtained using the method described in subsections \ref{sec:implementation} and \ref{sec:synthetic_spectra}. The insets show the subset with $\Delta {\rm AIC} < 0$. Left panel shows the low-energy indices $\alpha_1$ and $\alpha_2$ and the right panel shows the ratio of the peak energy to the break energy. The lines are smoothed kernels of each distribution. The excess of fits with $\alpha_2 \sim -2$ is due to bursts where the hardening of the spectrum is below the detector energy window (see main text). These spectra are not better described by 2SBPL function compared to a CPL function as evident from the inset. When the Lorentz factor increases, more of the curvature is visible in the detector and the excess disappears, which can be seen in the top middle panel of Figure \ref{fig:different_2SBPL}. Both panels are made with $\chi = 1$ and $z = 1$.}
    \label{fig:hist_2SBPL}
\end{figure*}

\subsection{Double smoothly broken power law}\label{sec:disc_2SBPL}
From Figure \ref{fig:typical_spectra} and the discussion in subsection \ref{sec:spectral_characteristics}, it is evident that the spectrum in the delayed acceleration scenario has more complexity than a simple CPL function. To probe this characteristic, we employ a double smoothly broken power-law (2SBPL) function as defined in \citet{Ravasio2018}. The 2SBPL is an empirical function with two low-energy power laws with indices $\alpha_1$ and $\alpha_2$, smoothly connected at a break energy $E_{\rm break}$. The low-energy part of the spectrum is smoothly connected to a high-energy power law with index $\beta$ at the peak energy $E_{\rm peak}^{\rm 2SBPL}$. The smoothness of the two breaks are determined by two smoothness parameters, which we fix to $n_1 = 5.38$ and $n_2 = 2.69$ similar to \citet{Ravasio2018} \citep[see also][]{Kaneko2006}. 

As in subsection \ref{sec:parameter_dist}, we fit 150 synthetic photospheric spectra with the 2SBPL function, with initial parameter distributions again given by Table \ref{tab:init_dist}. 
The distribution for the low low-energy spectral index, $\alpha_1$, and the high low-energy spectral index, $\alpha_2$, are shown in the left panel in Figure \ref{fig:hist_2SBPL}. 
In the right panel, we show the distribution for the ratio of peak energies to break energies.

By performing CPL fits to the same 150 input photospheric spectra and using the Akaike Information Criterion (AIC), we can evaluate which model is preferred. The AIC accounts for the number of free parameters in a model and a lower AIC value indicates a better model fit to the data. We find that $\sim 20$\% of cases have $\Delta {\rm AIC} \equiv {\rm AIC}_{\rm 2SBPL} - {\rm AIC}_{\rm CPL} < 0$ with $\sim 10$\% of total fits having $\Delta {\rm AIC} < -4$. As many of the generated synthetic spectra are well fitted with the CPL function within the GBM energy window, there is often no need for the extra degrees of freedom in the 2SBPL function and, thus, $\Delta {\rm AIC} > 0$ in a majority of cases. Still, this result shows that the 2SBPL function provides a better fit than the CPL function for a large fraction of photospheric spectra. The spectra that with $\Delta {\rm AIC} < 0$ have higher values of $\Gamma$ on average, leading to a $\sim 3$ times higher median value of the observed radiation luminosity, $L_{\gamma}$ (calculated using equation \eqref{eq:L_gamma} below), as compared to the whole fitted sample. Thus, we would expect the 2SBPL to be a better fit in bright bursts. 

The distributions of the parameter values obtained when fitting a 2SBPL function to KRA spectra have large dispersions. It is, however, remarkable that the obtained values are quite similar to those expected in marginally fast cooling synchrotron models ($\alpha_1 = -2/3,~\alpha_2 = -3/2,~E_{\rm peak}^{\rm 2SBPL}/E_{\rm break} \sim {\rm few}$). This similarity is especially clear in the case of larger Lorentz factors, as evident from the middle panels in Figure \ref{fig:different_2SBPL}. Obtaining values close to the values expected from synchrotron theory is not due to the KRA itself, i.e., there is no reason intrinsic to the model for why this would occur. Rather, it is an effect of the limited energy windows of the detectors and the curvature of the spectrum. If a given spectrum has more complexity than a single power law at low energies, then it is likely that the 2SBPL function will provide a better fit due to its higher flexibility. Furthermore, if the power-law index found when fitting a CPL function to the spectrum is $\alpha \sim -1$, then the two indices found in a 2SBPL fit are commonly going to satisfy $\alpha_2 < \alpha < \alpha_1$. Additionally, both the break energy and the peak energy are usually found within the limited detector energy window, leading to a moderate ratio between the two. The closeness of the two breaks can also be explained by the fact that the curvature of the 2SBPL function around $E_{\rm peak}^{\rm 2SBPL}$ cannot quite capture the smooth curvature around the peak in the input spectrum. Thus, $E_{\rm break} \lesssim E_{\rm peak}^{\rm 2SBPL}$ to account for the hardening below the peak. In many cases, therefore, $E_{\rm break}$ does not correspond to the break in X-rays, which is visible around $5~$keV in Figure \ref{fig:example_fit}. In Figure \ref{fig:hist_2SBPL}, these cases correspond to the overrepresentation of fits where $\alpha_2 \sim -2$, which corresponds to a flat power law in a $\nu F_\nu$ spectrum.

The above discussion illustrates that one has to be cautious when drawing conclusions regarding the physics based on fits of empirical functions alone, in particular over limited energy ranges. Identifying spectral indices close to $-0.67$ and $-1.5$ is not sufficient to conclude that the emission is synchrotron radiation. Such a spectrum can be produced by a dissipative photosphere, as evident from Figure \ref{fig:hist_2SBPL}, or by another emission mechanism that produces a curved spectrum within the energy range of the detectors. 

To break this model degeneracy, complementary and independent properties of the emission needs to be assessed. For instance, \citet{Ravasio2019} fitted a 2SBPL to the time-resolved data of the large, prompt flare in GRB 160821A and found averaged power law indices of $-0.74\pm0.15$ and $-1.51\pm0.17$. For this burst, further independent indications corroborate that the emission is indeed synchrotron. First, the flare has a large polarisation degree \citep[][although, see \citet{Lazzati2004}]{Sharma2019, Gill2020} and second, direct fits with an actual synchrotron model yield good agreement to the data and reveal an accelerated electron distribution consistent with the theory of particle acceleration in weakly magnetized flows \citep{Ryde2022arXiv}. Additional indirect arguments in favor of synchrotron origin for this burst is that the emission occurs late and has a long duration \citep{Li2021}.

Another way to break the degeneracy is to obtain early observations outside of the $\gamma$-ray band. A KRA spectrum is narrower than that predicted from synchrotron models. For instance, the asymptotic power-law slope at low energies in a broadened photospheric spectrum is much harder \citep[$\alpha \sim 0.4$,][]{Beloborodov2010, Acuner2019} than that expected from synchrotron emission \citep[$\alpha \sim -2/3$,][]{Sari1998}. Thus, optical observation have the power to differentiate between the models. Such a study was conducted in \citet{Oganesyan2019}. Unfortunately, current optical observations of GRBs always occur quite late (typically $\sim 100~$s after trigger). A few early optical observations may become available with the launch of the SVOM satellite in mid-2023 \citep{Wei2016}. Similarly, early high-energy data is a good way to distinguish between the models, as the KRA spectra drop rapidly in flux above the peak.

\subsection{Differences between short and long GRBs}

In this work, we have found that the low-energy index and the peak energy correlate with the injection radius, $r_0$. Furthermore, the observed variability time is also dependent on $r_0$. It is therefore interesting to speculate whether the difference in these properties between short and long GRBs could be due to differences in their initial collimation. 
\citet{Gottlieb2019,Gottlieb2021} found that in simulations of short GRBs, the collimation shock moves outwards and often breaks out of the surrounding neutron star ejecta. Hence, it may be difficult to prescribe a single value to $r_0$. However, if we for simplicity assume that $r_0$ is an order of magnitude smaller on average in short GRBs compared to long ones, then one can estimate (see equations \eqref{eq:E_peak_est} and \eqref{eq:tvar} below) that $E_{\rm peak}$ should be a factor 1.8 times higher and the variability time should be 10 times smaller in short GRBs compared to long ones. Moreover, by generating 150 spectra using $r_0 = 10^9$ (still with $\Gamma_0 = 4$) and fitting them with the CPL function, we find a mean value of $\langle \alpha\rangle = -0.55$, as compared to the mean value of $\langle \alpha\rangle = -0.76$ found for the distribution in Figure \ref{fig:hist_CPL} when $r_0 = 10^{10}~$cm. 

All these differences are close to the values observed. \citet{GolkhouButler2014} found the observed variability time to be a factor 12.5 smaller in short GRBs compared to long ones. In a time-integrated analysis of the first 10 years of \textit{Fermi} data, \citet{Poolakkil2021} found that $E_{\rm peak}$ was a factor of 2.6 higher in short GRBs compared to long ones on average. In the same work, \citet{Poolakkil2021} also found $\langle \alpha\rangle = -0.59\pm 0.5$ for short GRBs. This can be compared to the mean value $\langle \alpha\rangle = -0.80\pm 0.311$ found in the time-resolved analysis of \citet{Yu2016}, where 80 of the 81 bursts analysed were long GRBs.

\subsection{Additional predictions within the internal shock framework}\label{sec:additional_internal_predictions}
Within the internal shock framework employed, one can make many additional predictions regarding the emission. In this subsection, we give estimates of the expected peak energy, efficiency, isotropic gamma-ray luminosity, and observed variability time. However, we stress that since these predictions are based on the idealised internal shock framework, they should be evaluated with caution.
 
\textit{Peak energy.} In the downstream of the RMS, the photon distribution will interact with thermal electrons at temperature $\theta_{\rm C}$. In thermal Comptonization, the average relative energy gain per scattering is $\left<\Delta \epsilon / \epsilon \right> = 4\theta - \epsilon$. Thus, high-energy photons with $\epsilon > 4\theta$ lose their energy more quickly compared to the energy gained by low-energy photons. The energy loss halts once the energy approaches $4\theta_{\rm C}$, which is always greater than the mean photon energy of the distribution. Therefore, one can use $\bar{\epsilon}_d^{\rm ph}$ to get a lower-limit for the high-energy cutoff in the spectrum. Here, $\bar{\epsilon}_d^{\rm ph}$ is the average photon energy in the spectrum at the photosphere. Furthermore, as almost all spectra have a positive slope in their $\nu F_\nu$ spectrum (see subsection \ref{sec:spectral_characteristics}), $\bar{\epsilon}_d^{\rm ph}$ should be a good lower limit for the observed peak energy. The dissipated energy per proton in the RMS is approximately $(\gamma_u - 1)m_p c^2$. Hence, the average photon energy at the photosphere can be estimated as


\begin{equation}\label{eq:epsilon_d_ph}
    \bar{\epsilon}_d^{\rm ph} \approx \frac{(\gamma_u - 1)m_p}{\tilde{n} m_e}\frac{6}{5}\tau_i^{-2/3}.
\end{equation}

\noindent where the term $(6/5)\tau_i^{-2/3}$ accounts for the adiabatic cooling of the distribution. With a Doppler factor of $5\Gamma/3$, accounting for redshift, and using equations \eqref{eq:gamma_u} and \eqref{eq:tau}, one gets an estimated lower limit for the peak energy

\begin{equation}\label{eq:E_peak_est}
\begin{split}
    E_{\rm peak}^{\rm ll} &\approx \frac{5.83}{3} \frac{\Gamma_s(\sqrt{\psi} - 1)^2}{\tilde{n}(1+z)} m_p c^2 \tau_i^{-2/3} \\[2mm]
    & = 31 \, {\rm keV} ~ \chi^{1/4} \, r_{0,10}^{-1/4} \, \beta_0^{-1/4} \, \Gamma_0^{1/2} \, \Gamma_{s,1.5}^{5/4}\, \tau_{s,1.5}^{-5/12} \psi_1,
\end{split}
\end{equation}

\noindent where the numerical factor 5.83/3 transforms the average photon energy to the peak energy in a $\nu F_\nu$ spectrum \citep{Ryde2017}, we used $(\gamma_u-1)\approx (\sqrt{\psi} - 1)^2/2\sqrt{\psi}$ obtained from equation \eqref{eq:gamma_u}, we used equation \eqref{eq:tau}, and we only show the leading parameter dependence of $\psi$ in the final line. The estimate above explains why the fitted values for $E_{\rm peak}$ are higher on average when $r_0$ is $10^7~$cm as compared to $10^{10}~$cm (see Figure \ref{fig:hist_CPL}).

We compared $E_{\rm peak}^{\rm ll}$ to the values of the peak energy found for the fits in the histograms in Figure \ref{fig:hist_CPL}. In the early acceleration scenario, the values of $E_{\rm peak}$ are very close to $E_{\rm peak}^{\rm ll}$ indicating almost complete thermalization of the high-energy photons. In the delayed acceleration scenario, $E_{\rm peak}^{\rm ll}$ is commonly a factor of a few lower than the peak energy, indicating incomplete thermalization. For instance, the fit shown in Figure \ref{fig:example_fit} yielded $E_{\rm peak} = 150~$keV compared to $\sim 62~$keV obtained using equation \eqref{eq:E_peak_est}. 


\textit{Efficiency.} As explained in Appendix \ref{App:internal_collision}, both fireballs are assumed to have similar total masses. With the mass in each of the two fireballs denoted by $M = N_p m_p$, the total energy emitted from the central engine is $E_{\rm tot} = (\psi + 1)\Gamma_s N_p m_p c^2$. The total radiated energy in the central engine frame is $E_\gamma \approx 2(5\Gamma/3) N_\gamma \, \bar{\epsilon}_d^{\rm ph} m_e c^2$ where $N_\gamma$ is the total number of photons in each fireball, $5\Gamma/3$ is the Doppler boost, and an additional factor 2 comes from the fact that there are two fireballs. With $N_\gamma = N_p \tilde{n}$, we get an estimated radiation efficiency of

\begin{equation}\label{eq:efficiency}
    \eta = \frac{E_\gamma}{E_{\rm tot}} = \frac{10 \sqrt{\psi} \, \tilde{n} \, \bar{\epsilon}_d^{\rm ph}}{3(\psi + 1)}\frac{m_e}{m_p} \approx \frac{2(\sqrt{\psi}-1)^2}{(\psi + 1)}\tau_i^{-2/3},
\end{equation}

\noindent where we have inserted $\bar{\epsilon}_d^{\rm ph}$ from equation \eqref{eq:epsilon_d_ph}. Using the more accurate value of $\bar{\epsilon}_d^{\rm ph}$ determined from the shock jump conditions, the mean efficiency is $\sim 6$\%, which agrees quite well with the estimate in equation \eqref{eq:efficiency}. As evident from the equation above, the efficiency is almost entirely determined by the optical depth of the RMS, meaning that photospheric emission can be very efficient if the dissipation occurs close to the photosphere.

\textit{Isotropic gamma-ray luminosity.} The total proton number is the same in both fireballs. In the slow fireball, it can be estimated as $N_p = E_s/\Gamma_s m_p c^2 \approx L_s (l_s/c)/\Gamma_s m_p c^2$, where $E_s$ it the total emitted energy in the slow fireball and $l_s$ is the observed width of the slow fireball. The RMS traverses the upstream in a doubling of the radius, which implies that $l_s \sim \delta t \, c = \chi r_0$. The source frame $\gamma$-ray luminosity can be estimated as $L_\gamma \sim E_\gamma/(t_{\rm var}/(1+z))$, where $t_{\rm var}$ is the observed variability time, which divided by $(1+z)$ is the source variability time. Thus,

\begin{equation}\label{eq:L_gamma}
\begin{split}
    L_\gamma &\sim \frac{E_\gamma(1+z)}{t_{\rm var}} \approx \frac{2(\sqrt{\psi}-1)^2\psi L_s}{\tau_i^{5/3}} \\[2.2mm]
    &= 1.6\times 10^{51} \, {\rm erg}~ {\rm s}^{-1} ~ \chi \, r_{0,10}\, \Gamma_{s, 1.5}^5 \, \tau_{s, 1.5}^{-2/3} \psi_1^2,
\end{split}
\end{equation}

\noindent where we inserted $L_s$ and $t_{\rm var}$ from equations \eqref{eq:L_s} and \eqref{eq:tvar} and only show the leading dependence on $\psi$ in the last line. 

\textit{Variability time.}
In our delayed acceleration scenario, the values for the photospheric radius obtained in this work are of the order $\sim 10^{15}~$cm, which is much larger than what is commonly discussed for photospheric models ($\sim 10^{11}{\rm -}10^{12}~$cm). The large radii are due in part to the high isotropic equivalent luminosities required to keep the shocks subphotospheric ($\sim 10^{54}~$erg$~$s$^{-1}$) and in part due to the low Lorentz factors ($\sim 100$). These two factors alone increase the photospheric radius to $6\times 10^{14}~$cm \citep[e.g.,][]{Peer2015}. Additionally, the compression in the downstream of the RMS further pushes the photospheric radius outwards by a factor of a few. 

The large photospheric radius directly solves one problem with photospheric models in GRBs, which is the short variability times predicted. For an instantaneous flash in a thin shell occurring at a radius $r$, the emitted light reaches the observer over a timescale $t_{\rm var} = (1+z)r/2\Gamma^2c$. With a canonical photospheric radius at $\sim 10^{12}~$cm and Lorentz factor of $\Gamma = 100$, one expects millisecond variability, compared to the commonly observed (minimum) variability time of $\sim 2.5 \,$s in long GRBs \citep{GolkhouButler2014}. However, in the delayed acceleration scenario, one gets

\begin{equation}\label{eq:tvar}
\begin{split}
    t_{\rm var} &= (1+z) \frac{r_{\rm ph}}{2\Gamma^2 c} = (1+z) \frac{\chi r_0 \tau_i}{\psi c} \\[2mm]
    &= 6.3 \, {\rm s} ~ \chi \, r_{0, 10} \, \tau_{s,1.5} \, \psi_1^{-1},
\end{split}
\end{equation}

\noindent where we used equation \eqref{eq:tau} and set $z = 1$. Note that in a realistic scenario with multiple collision below the photosphere, variability on different timescales can be observed simultaneously in the light curve.

In \citet{ReesMeszaros2005} it was suggested that $\chi = \theta_j$, where $\theta_j$ is the half-opening angle of the jet. With $\theta_j = 5^\circ$, the luminosity, photospheric radius, and variability time would all decrease by a bit more than an order of magnitude. 

\section{Summary and conclusion}\label{sec:Summary}
We have studied what observational characteristics are expected in dissipative photospheric models of GRBs, when the dissipation is in the form of an RMS. The RMS was modeled using the KRA developed in \citet{Samuelsson2022}, which allowed us to investigate the spectral properties quantitatively due to its low computational cost. To estimate relevant shock initial conditions in the context of GRBs, we employed a simple internal collision framework (Figure \ref{fig:collision_schematic}). We found that when the free fireball acceleration starts close to the central engine, the spectral slope was hard. However, in hydrodynamical jet simulations, one commonly finds that the acceleration is delayed due to collimation, resulting from high-pressure material surrounding the jet \citep{Lazzati2009, Lopez-Camara2013, Gottlieb2018, Gottlieb2019, Gottlieb2021}. In this case, the obtained spectrum was softer and exhibited an additional break in X-rays (Figure \ref{fig:typical_spectra}). The spectrum was well fitted over the limited GBM energy range with a phenomenological CPL function (Figure \ref{fig:example_fit}).

We continued by generating 150 synthetic photospheric spectra as they would appear in the \textit{Fermi} GBM detector and fitting them with the CPL function. The derived histograms for the low-energy photon index, $\alpha$, and peak energy, $E_{\rm peak}$, were compared to the catalogued distributions found in a time-resolved analysis of bright \textit{Fermi} burst \citep{Yu2016}. We found that when the free fireball acceleration started close to the central engine, the spectra were generally hard with $\alpha\gtrsim -0.75$, inconsistent with a majority of observed GRBs. 
In the scenario where the acceleration was delayed, the average value for the power-law index was $\alpha \sim -0.8$ with a range of values consistent with the dominant fraction of catalogued bursts. This showed that photospheric emission with correctly modeled prior dissipation can be much softer than what is commonly argued. The obtained values of $E_{\rm peak}$ were similarly in general agreement with the observations (Figure \ref{fig:hist_CPL}). 

We also fitted 150 synthetic spectra with a 2SBPL function and found that such a fit was strongly preferred over the CPL function in $\sim 10\%$ of cases. The bursts where the 2SBPL had lower AIC value than the CPL function had a $\sim 3$ times higher median gamma-ray luminosity compared to the sample as a whole. The average values obtained for the low-energy indices $\alpha_1$ and $\alpha_2$, as well as the ratio of peak energy to break energy are quite similar to those expected in marginally fast cooling synchrotron emission models (Figure \ref{fig:hist_2SBPL}). As this is not an intrinsic property of our dissipative photospheric model, we argued that this is an effect of the limited energy windows of the detectors and the smooth curvature of the spectrum. 
Thus, one must be cautious when drawing conclusions about the underlying physics based on empirical spectral fits, as different physical models can yield similar fitted parameter values. We listed several ways this degeneracy could be broken, including polarization measurements, fitting physical models directly to the data, and early optical and high-energy observations. 


Finally, we found that for a large range of optical depths, the radiation in the RMS does not have time to reach steady state before the dissipation ends. We called these shocks ``optically shallow shocks'' and found in equation \eqref{eq:tau_ss2} that they occur for optical depths as deep as $\tau \sim 55 \, u_u^{-2}$. Optically shallow shocks are interesting, as signatures of the shock formation may still be imprinted in the observed spectrum.

In conclusion, we find that photospheric emission that has been energized by a subphotospheric internal collision is spectrally consistent with the majority of observed GRBs. Additionally, this type of photospheric emission can explain various observational features such as an additional break in X-rays, the observed variability time, and possibly some of the main differences between short and long GRBs. This requires that the free fireball acceleration is delayed until $\sim 10^{10}~$cm, primarily to increase the photon-to-proton ratio (see also Table \ref{tab:parameter_dist}). The main caveats relate to the internal collision framework used and are the low efficiency ($\sim 6$\%) and the high isotropic equivalent luminosity required ($L \sim 10^{54}~$erg$~$s$^{-1}$). The luminosity requirement can be relaxed if the photon-to-proton ratio can be otherwise increased, e.g., by additional photon production at a collisionless subshock.\\[1.5mm]


We thank the anonymous reviewer for many useful comments. We acknowledge support from the Swedish National Space Agency (196/16) and the Swedish Research Council (Vetenskapsr\aa det, 2018-03513 and 2022-00347). We thank Dr. Christoffer Lundman and Dr. Damien B\'egu\'e for fruitful discussions. 
F.R. is supported by the G\"oran Gustafsson Foundation for Research in Natural Sciences and Medicine. This research made use of the High Energy Astrophysics Science Archive Research Center (HEASARC) Online Service at the NASA/Goddard Space Flight Center (GSFC). In particular, we thank the GBM team for providing the tools and data.

\bibliography{References}{}
\bibliographystyle{aasjournal}

\appendix

\section{Extending the KRA to treat shock formation}\label{App:theta_r_v}

In the collision of two blobs, the central region between the blobs is initially compressed adiabatically. The compression increases the temperature and, in the opaque systems below the photosphere, the radiation pressure. When the downstream radiation pressure becomes comparable to the ram pressure of the incoming baryons, which roughly occurs when the average photon energy in the compressed region equals $\bar{\epsilon}_d$, two shocks form \citep{Beloborodov2017}. Hence, the shocks initially contains a thermal distribution of photons with average energy $\bar{\epsilon}_d$. 

The shock jump conditions dictate that the average energy of photons advected into the downstream should be $\bar{\epsilon}_d$ (ignoring adiabatic cooling, which is small over the lifetime of the of the shock and does not change the reasoning). 
Since the escape probability from the RMS to the downstream is energy independent, the average energy in the RMS zone must also equal $\bar{\epsilon}_d$.
However, this requires $\langle \Delta \epsilon / \epsilon \rangle$ to vary during the shock formation. Thus, in the KRA we want a varying effective electron temperature, $\theta_r^{v}$, that keeps $\bar{\epsilon}_r$ constant (which in turn keeps $\bar{\epsilon}_d$ constant) and which equals $\theta_r$ as given by equation \eqref{eq:4theta_r} when the radiation in the RMS reaches steady state.

\begin{figure}
    \centering
    \includegraphics[width=\columnwidth]{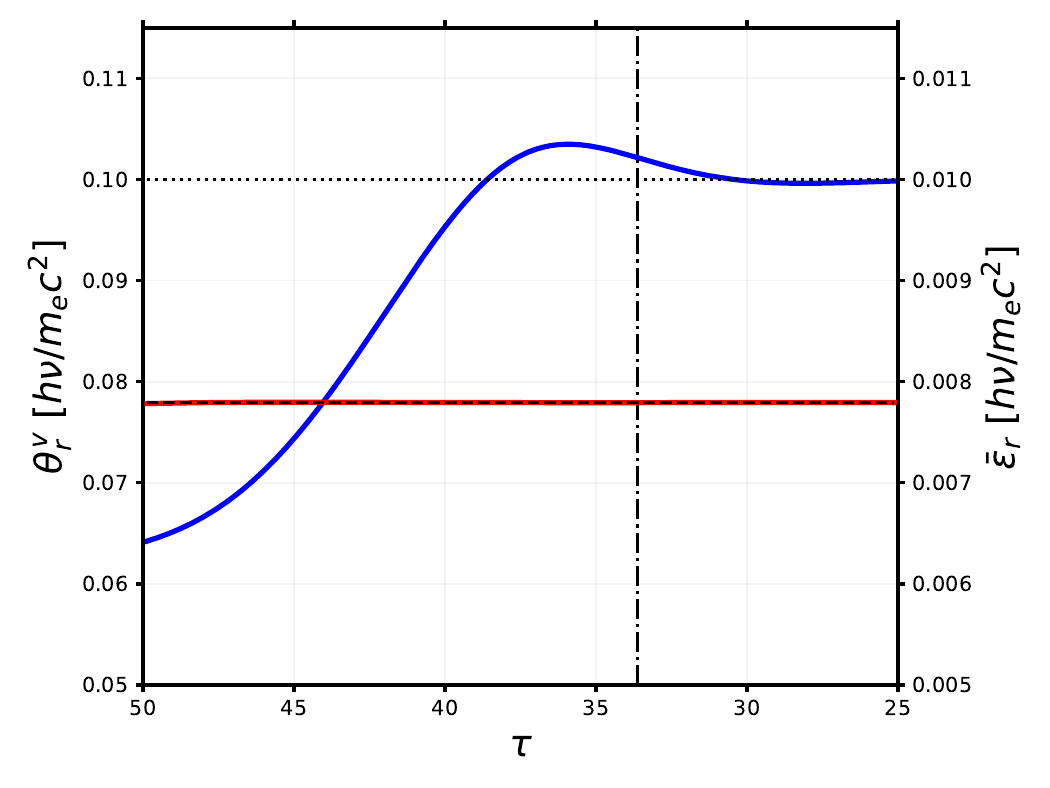}
    \caption{Figure showing $\theta_r^v$ (left axis, blue line) and $\bar{\epsilon}_r$ (right axis, red line) as a function of optical depth for a simulation. The effective electron temperature $\theta_r^v$ initially grows as the Compton temperature in the shock grows. It tends to its steady state value $\theta_r$, which is indicated by the horizontal dotted line. As expected, the average photon energy in the RMS, $\bar{\epsilon}_r$, remains at all times constant and equal to $\bar{\epsilon}_d$, which is indicated by the horizontal dashed line. The vertical dot-dashed line shows our estimate of $\tau_{\rm ss}$ calculated using equation \eqref{eq:tau_ss3}. It slightly underestimates when equilibrium is reached, which is expected since equation \eqref{eq:tau_ss3} assumes a constant energy gain per scattering of $4\theta_r$, while the true energy gain is equal to $\theta_r^v$, which is lower than $\theta_r$ initially. Adiabatic cooling is not included. Parameter values are $\tau\theta = 5$, $\theta_r = 0.1$, $R = 500$, $y_r = 1.5$. }
    \label{fig:theta_r}
\end{figure}

The average photon energy in the RMS zone can change due to Compton scatterings with the hot electrons and photons entering and leaving the shock region. The fraction of photons that leave the shock between an optical depth of $\tau$ and $\tau + d\tau$ is given by $(4\theta_r/y_r) \, d\tau$ when the radiation is in steady state \citep[see][for details]{Samuelsson2022}. Assuming the number of photons that pass through the shock is the same during the shock formation as during the later steady state phases, this applies to the early stages as well. The expression for the average energy change in the RMS zone in the KRA is then given by 
\begin{equation}\label{eq:deps_r_dtau}
    \frac{d\bar{\epsilon}_r}{d\tau} = 4\bar{\epsilon}_r(\theta_r^v - \theta_{{\rm C},r}) + \frac{4\theta_r}{y_r}(\bar{\epsilon}_u - \bar{\epsilon}_r),
\end{equation}
\noindent where $\theta_{{\rm C},r}$ is the Compton temperature of the photons in the RMS zone. Using the equation above, we wish to find the shock temperature $\theta_r^v$ that satisfies $\frac{d\bar{\epsilon}_r}{d\tau} = 0$. To ensure stability, we replace $\bar{\epsilon}_r$ with $\bar{\epsilon}_d$ in the last term in the last parenthesis, since this is the average photon energy that should be injected into the downstream. Note that the average energy of the photons injected into the downstream in the simulation is still $\bar{\epsilon}_r$; this prescription is only used in the calculation of $\theta_r^v$. Furthermore, with a fine enough radial resolution, we checked that this choice has no effect since at all times $\bar{\epsilon}_r = \bar{\epsilon}_d$. 
Solving for $\theta_r^v$, one gets
\begin{equation}
    \theta_r^v = \theta_{{\rm C},r} + \frac{\theta_r}{y_r}\left(\frac{\bar{\epsilon}_d}{\bar{\epsilon}_r} - \frac{\bar{\epsilon}_u}{\bar{\epsilon}_r} \right).
\end{equation}
In Figure \ref{fig:theta_r}, we show how $\theta_r^v$ varies during the shock formation. Initially, it is lower than the asymptotic value $\theta_r$. 
As photons are heated, the photon distribution changes and $\theta_r^v$ grows. In terms of a true RMS, this can be thought of as a steepening of the velocity gradient when the shock is initially launched. The growth saturates when the influx of cold photons starts to dominate over the energy increase due to scatterings, resulting in a decrease of the Compton temperature, and, hence, of $\theta_r^v$. Finally, $\theta_r^v$ converges to $\theta_r$. The qualitative behavior is independent on the choice of initial parameters. 

\begin{figure*}
    \centering
    \includegraphics[width=\columnwidth]{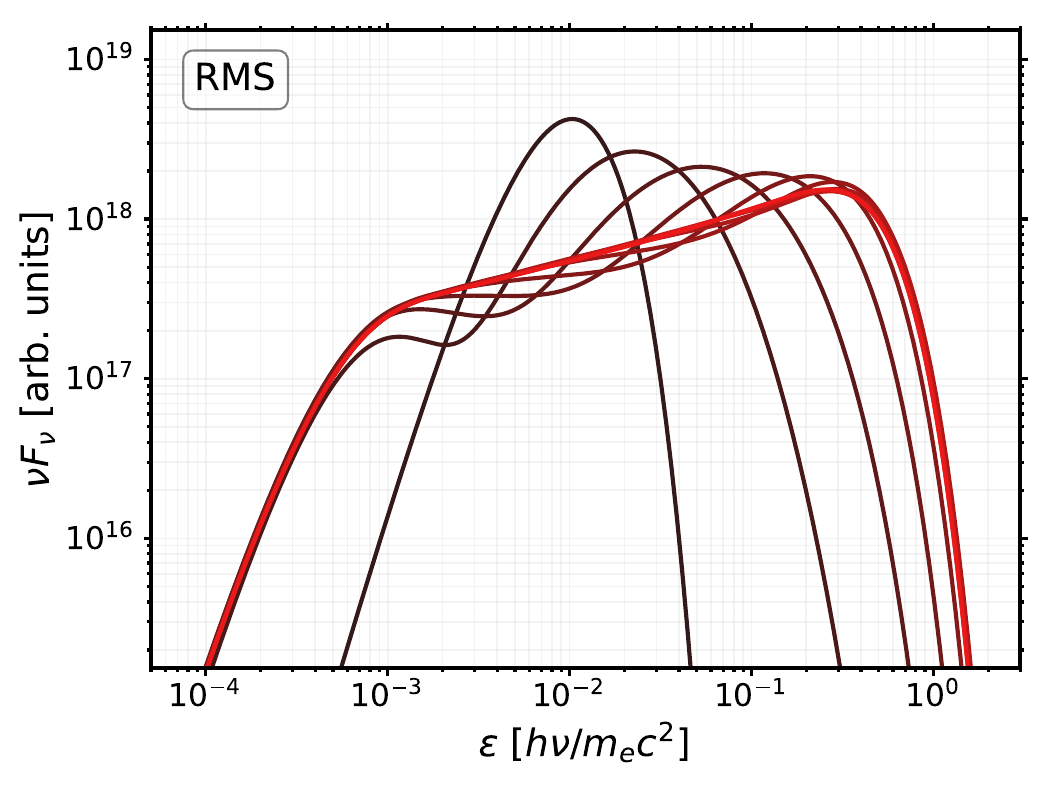}
    \includegraphics[width=\columnwidth]{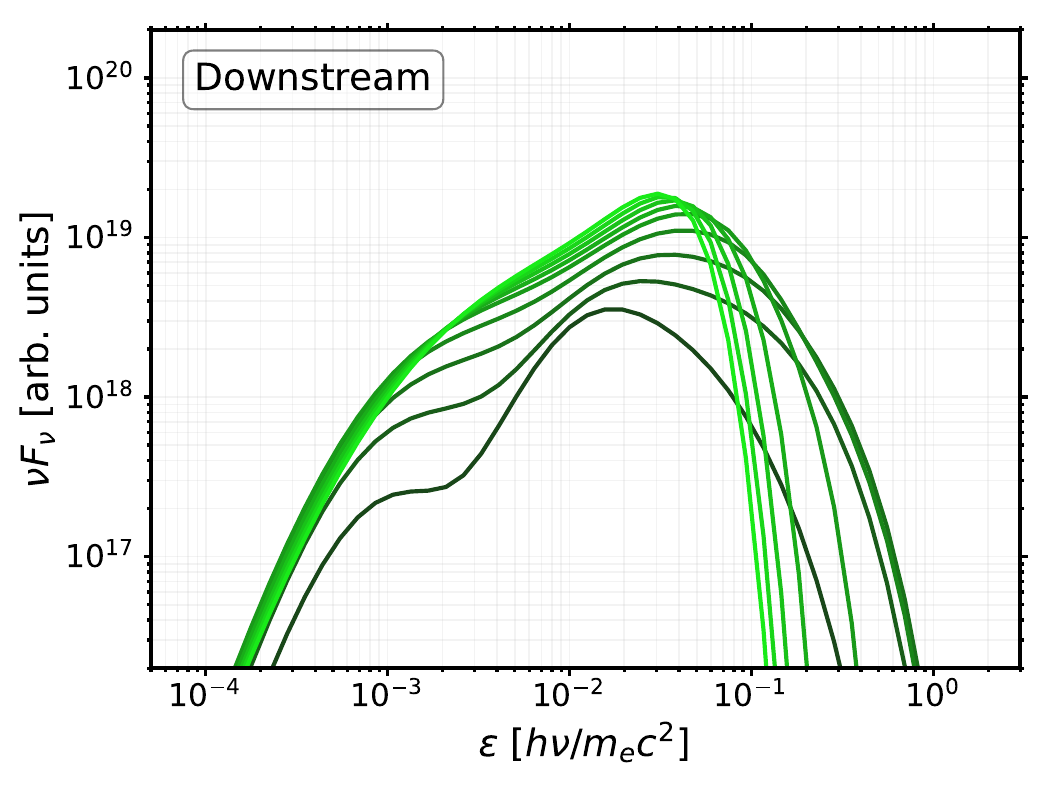}
    \caption{\textit{Left}: Evolution of the RMS spectrum in a simulation. Ten snapshot spectra are shown, taken at even intervals of $\tau$ ranging from the initial value $\tau = 50$ (black) until the shock has crossed the upstream at $\tau = 25$ (red). Initially, the shock contains a Wien thermal photon distribution with mean energy $\bar{\epsilon}_d$. As the shock evolves, the initial peak moves to higher energies and broadens, while new cold photons enter the shock from the upstream. At $\tau\approx 33$ (snapshot seven), steady state has been reached and the final spectra are overlapping. This corresponds well with the estimate shown in Figure \ref{fig:theta_r}. \textit{Right}: Similar to the left panel but for the downstream spectrum. This time the spectra range from $\tau = 50$ all the way up until the photosphere at $\tau = 1$ (from darker to brighter green). Note that the spectrum for $\tau = 50$ is not visible since the downstream does not contain any photons at the start of the simulation. For this specific choice of parameters, the optical depth was large enough to erase all signatures of the shock formation in the final spectrum. Adiabatic cooling and the broadening effects explained in Appendix \ref{App:broadening} are not included. Parameter values are the same as in Figure \ref{fig:theta_r}.}
    \label{fig:shock_formation}
\end{figure*}

In Figure \ref{fig:shock_formation}, we show how the RMS and the downstream photon distribution evolves in a simulation by plotting ten snapshot spectra of each. From the Figure, one can see how the RMS spectrum appears before it reaches steady state. The downstream initially shows clear signs of the shock formation, but for this simulation they are washed out before the downstream reaches the photosphere. Indeed, this is the case for most of the spectra in Figure \ref{fig:hist_CPL}: less than 10\% of the spectra in Figure \ref{fig:hist_CPL} qualify as optically shallow shocks, as calculated by equation \eqref{eq:tau_ss4}.

\section{Details regarding the subphotospheric internal collision framework}\label{App:internal_collision}
We consider a variable central engine that launches a slower and a faster fireball with a time separation $\delta t$. The two fireballs are assumed to have similar comoving densities, $\rho$, and similar total masses, $M$, but different isotropic equivalent luminosities, $L_s$ and $L_f$, which results in different terminal Lorentz factors, $\Gamma_s$ and $\Gamma_f$. For a collision to occur, we require $\Gamma_f/\Gamma_s \equiv \psi > 1$. That the two fireballs have similar proper densities implies that the ratio of the isotropic equivalent luminosities of the two fireballs satisfies $L_f/L_s = \psi^2$. 

Initially, the fireballs accelerate until most internal energy has been converted into bulk kinetic energy. This occurs at the saturation radius 

\begin{equation}
	r_{{\rm sat}, i} = \frac{r_0\Gamma_i}{\Gamma_0},
\end{equation}

\noindent where the subscript $i$ indicates fireball $i$. Above, $r_0$ and $\Gamma_0$ are the radius from the central engine and the Lorentz factor at the base of the conical outflow. These could in principle be slightly different for the two fireballs but here we assume that they are the same. 
The radius at which the two shells collide is given by

\begin{equation}\label{eq:rcoll}
\begin{split}
	r_{\rm coll} \approx 2 \Gamma_s^2 c \delta t = 2 \Gamma_s^2 \chi r_0 = 2.0 \times 10^{13} \, {\rm cm} ~ \chi \, r_{0,10} \, \Gamma_{s, 1.5}^2,
\end{split}
\end{equation}

\noindent where we have parameterized $\delta t = \chi r_0/c$, that is, we assume that the time separation between the two fireballs is related to the light crossing time at the radius where the free acceleration begins. In this work, we use $\chi =1$ in both the early and delayed acceleration scenario. However, we include it in the equations to show the parameter dependence.

The collision will be subphotospheric if $r_{\rm coll}$ is smaller than the photospheric radius of the slower shell.
In a conical outflow where the density drops as $r^{-2}$, the optical depth toward the observer at the collision radius for a fluid element in the slower shell is given by $\tau_s = r_{{\rm ph},s}/r_{\rm coll}$. As we are only interested in collisions that occur below the photosphere, we use $\tau_s$ as a free parameter in our collision model. 
Apart from $\tau_s$, we use $\psi$ and $\Gamma_s$ as free parameters. 

The photospheric radius for the slower fireball with our parameterization is given by

\begin{equation}\label{eq:rphs}
\begin{split}
	r_{{\rm ph},s} = r_{\rm coll} \, \tau_s \approx 6.3\times 10^{14} \, {\rm cm} ~ \chi \, r_{0,10} \, \Gamma_{s, 1.5}^2 \, \tau_{s, 1.5}.
\end{split}
\end{equation}

\noindent The photospheric radius can also be written as $r_{{\rm ph},s} = L_s \sigma_{\rm T}/(8\pi m_p c^3 \Gamma_s^3)$.
Solving for the isotropic luminosity of the slower fireball one gets

\begin{equation}\label{eq:L_s}
\begin{split}
    L_s &= \frac{16 \pi m_p c^3 \chi r_0 \, \Gamma_s^5 \, \tau_s}{\sigma_{\rm T}} \\
    &= 3.4\times 10^{52} \, {\rm erg}~ {\rm s}^{-1} ~ \chi \, r_{0,10}\, \Gamma_{s, 1.5}^5 \, \tau_{s, 1.5}.
\end{split}
\end{equation}

The comoving temperature of the plasma in fireball $i$ at the collision radius, assuming the collision occurs above the saturation radii of both fireballs, is given by

\begin{equation}\label{eq:theta_u1}
	\theta_{u,i} = \theta_{0,i} \, \frac{\Gamma_0}{\Gamma_i}\left(\frac{r_{\rm coll}}{r_{{\rm sat},i}}\right)^{-2/3},
\end{equation}

\noindent where $\theta_{0, i}$ is the temperature of fireball $i$ at the base of the conical outflow. Assuming a thermodynamic equilibrium at the base, the initial temperature can be found through conservation of momentum flux as 

\begin{equation}
	\theta_{0,i} = \frac{k}{m_e c^2}\left(\frac{L_i}{16\pi r_0^2 c a \, \Gamma_0^2 \beta_0}\right)^{1/4},
\end{equation}

\noindent where $k$ is the Boltzmann constant, $a$ is the radiation constant, and $\beta_0 = \sqrt{1-1/\Gamma_0^2}$. Inserting this back into equation \eqref{eq:theta_u1} and using equation \eqref{eq:L_s}, one gets

\begin{equation}\label{eq:theta_u2}
\begin{split}
	\theta_{u,s} &= \frac{k\Gamma_0}{m_e c^2\Gamma_s} \left(\frac{m_p c^2 \chi \Gamma_s^5 \tau_s}{\sigma_{\rm T} a r_0 \Gamma_0^2 \beta_0} \right)^{1/4} \left(2\Gamma_s\chi \Gamma_0 \right)^{-2/3}\\[2mm]
	&=1.4\times 10^{-4}~ \chi^{-5/12} \, r_{0,10}^{-1/4} \, \beta_0^{-1/4} \, \Gamma_0^{-1/6} \, \Gamma_{s,1.5}^{-5/12}\, \tau_{s,1.5}^{1/4} ,\\[2mm]
	\theta_{u,f} & = \psi^{1/6} \, \theta_{u,s}. 
\end{split}
\end{equation}

\noindent The last line in the equation above indicates that the value and parameter dependence for $\theta_{u,f}$ is the same as that for $\theta_{u,f}$, apart from a factor $\psi^{1/6}$.

The photon-to-proton ratio, $\tilde{n}$, can be estimated assuming that photons and pairs are in thermodynamic equilibrium at the base of the jet, that all pairs have recombined before the collision radius, and that no new photons have been produced. In that case, one gets \citep{Bromberg2011b, Levinson2012}

\begin{equation}\label{eq:tilde_n}
\begin{split}
	\tilde{n}_s &= \frac{\Gamma_s}{4\theta_{0,s}\Gamma_0}\frac{m_p}{m_e} = \frac{m_p c^2}{4k} \left(\frac{\sigma_{\rm T} a \beta_0 r_0}{m_p c^2 \Gamma_0^2 \Gamma_s \chi \, \tau_s} \right)^{1/4}
	\\[2mm]
	& = 2.5 \times 10^5 ~ \chi^{-1/4} \, r_{0,10}^{1/4} \, \beta_0^{1/4} \, \Gamma_0^{-1/2} \, \Gamma_{s,1.5}^{-1/4}\, \tau_{s,1.5}^{-1/4}, \\[2mm]
	\tilde{n}_f & = \psi^{1/2} \, \tilde{n}_s. 
\end{split}
\end{equation}

\noindent At the collision radius, the adiabatic compression between the fireballs launches a forward shock that propagates into the slower fireball and a reverse shock that propagates into the faster fireball. From equations \eqref{eq:theta_u2} and \eqref{eq:tilde_n}, we see that the upstream conditions for these two shocks are slightly different in the two fireballs. However, these differences may easily be outweighed by the uncertainties in $r_0$ and $\Gamma_0$. For simplicity, in this work we assume the upstream conditions in both shocks are similar. For the results presented in Section \ref{sec:Results}, we used the values obtained for the forward shock, i.e., the upstream conditions in the slow fireball. However, in the main text, the subscript $s$ has been dropped. Using the upstream conditions in the faster fireball instead, the photon-to-proton ratio increases, which in turn softens the power-law and leads to to lower values of both $\alpha$ and $E_{\rm peak}$. The results obtained when using $\theta_{u,f}$ and ${\tilde n}_f$ are shown in Appendix \ref{App:different_params}.

The Lorentz factor of the causally connected downstream in between the two shocks, given similar proper densities of the two fireballs, is given by \citep{Kobayashi1997, Samuelsson2022}

\begin{equation}\label{eq:Gamma}
    \Gamma = \sqrt{\Gamma_s\Gamma_f} = \Gamma_s \sqrt{\psi} = 100 ~ \Gamma_{s,1.5} \, \psi_{1}^{1/2}.
\end{equation}

\noindent The assumption of similar proper densities in the two shells assures that the upstream relativistic velocity, $u_u$, is the same for the reverse and the forward shock. To determine the upstream relativistic velocity one must solve the shock jump conditions, which is done for all the results presented in main text. However, it is useful with analytical estimates, which we give below. The Lorentz factor of the incoming upstream can be estimated by

\begin{equation}\label{eq:gamma_u}
    \gamma_u \approx \frac{\Gamma_{\rm fs}}{2\Gamma_s} + \frac{\Gamma_s}{2\Gamma_{\rm fs}} \approx \frac{\psi + 1}{2\sqrt{\psi}},
\end{equation}

\noindent where $\Gamma_{\rm fs}$ is the Lorentz factor of the forward shock as measured in the lab frame, and we used $\Gamma_{\rm fs} \approx \Gamma$ for the last approximation. This gives

\begin{equation}
    u_u \approx \frac{\psi - 1}{2\sqrt{\psi}} \equiv f(\psi)\sqrt{\psi},
\end{equation}

\noindent where $f(\psi) = \frac{1}{2}(1 - 1/\psi)$ satisfies $1/4 < f(\psi) < 1/2$, as long as $\psi > 2$. This estimate shows that the dimensionless momentum is expected to be mildly relativistic with $u_u \lesssim 3$ for a wide range of $\psi$.

Although the Lorentz factor of the downstream is larger than the Lorentz factor of the slower fireball, the optical depth of a downstream fluid element toward the observer is actually slightly larger due to compression of the plasma \citep{LevinsonNakar2020}. (This is only true if one ignores diffusion out of the downstream. Indeed, diffusion in a thin shell might make photons decouple from the downstream before reaching the photosphere \citep{Ruffini2013, Begue2013}. Therefore, an implicit assumption of our model is that the specific collision considered is embedded in a jet where many such collisions occur. We note, however, that the diffusion time becomes comparable to the dynamical time close to the photosphere and that all our shocks have finished dissipation before this happens. Indeed, the shape of the spectrum is determined at early times before diffusion becomes relevant, as the number of scatterings decreases with density. Including escape via diffusion would only slightly decrease the amount of adiabatic cooling of the spectrum.) The optical depth for the downstream at the collision radius is given by 

\begin{equation}\label{eq:tau}
    \tau_i = \frac{u_u}{u_d}\frac{\tau_s}{\sqrt{\psi}} \equiv \tilde{f} \tau_s,
\end{equation}

\noindent where the compression is given by the ratio $u_u/u_d$, coming from the conservation of baryon number, and the factor $1/\sqrt{\psi}$ accounts for the increased Doppler boost. The function $\tilde{f}$ is monotonically decreasing with $\psi$ and depends only weakly on the other parameters. Its numerical value is $\tilde{f} \approx 4$ for $\psi = 2$ and $\tilde{f} \approx 2.6$ for $\psi = 30$. The compression also implies that the photospheric radius moves outwards by a factor $\tilde{f}$, as compared to the photospheric radius of the slower fireball. By looking at equation \eqref{eq:rphs}, we see that the photosphere can be as far out at $r_{\rm ph} \gtrsim 10^{15}~$cm. Some implications of this is discussed in subsection \ref{sec:additional_internal_predictions}.

We continue with analytical estimates of the KRA parameters. The upstream temperature is given by

\begin{equation}\label{eq:theta_u_K2}
    \theta_{u, {\rm K}} = (\tilde{f}\sqrt{\psi})^{1/3} \, \theta_u,
\end{equation}

\noindent where $\theta_u$ is given in \eqref{eq:theta_u2}. The steady state value of the effective electron temperature can be estimated as

\begin{equation}
    \theta_{r} \approx \frac{f^2\psi \ln(\bar{\epsilon}_d/\bar{\epsilon}_u)}{4\xi} = 0.042\, \psi_1 \ln(\bar{\epsilon}_d/\bar{\epsilon}_u)_2,
\end{equation}

\noindent where we used $\bar{\epsilon}_d/\bar{\epsilon}_u = 100$ in the last equality, neglecting their parameter dependence for simplicity as it only enters logarithmically and therefore has a small effect. The ratio $R = \theta_r/\theta_{u,{\rm K}}$ is then estimated as

\begin{equation}\label{eq:R}
\begin{split}    
    R \approx 140 ~ \chi^{5/12} \, r_{0,10}^{1/4} \, \beta_0^{1/4} \, \Gamma_0^{1/6} \, \Gamma_{s,1.5}^{5/12}\, \tau_{s,1.5}^{-1/4} \, \psi_1^{5/6} \,  \ln(\bar{\epsilon}_d/\bar{\epsilon}_u)_2,
\end{split}
\end{equation}

\noindent and $\tau \theta = \tau_i \theta_r$ is given by 

\begin{equation}
    \tau\theta \approx \frac{f^2\psi \ln(\bar{\epsilon}_d/\bar{\epsilon}_u)\, \tilde{f}\tau_s}{4\xi} = 4.0 ~ \tau_{s,1.5} \, \psi_1 \ln(\bar{\epsilon}_d/\bar{\epsilon}_u)_2.
\end{equation}

\noindent Although we cannot estimate the parameter dependence on $y_r$ analytically, its numerical value is $y_r \approx 1.6$ for the values used in this subsection.

\section{High-latitude and radial contributions to the photospheric emission}\label{App:broadening}

Three ingredients are needed in order to post-process the transition to transparency: the local comoving spectral shape at the photosphere, the distribution of the last scattering radius and angle, and the radial cooling of radiation as the jet transitions to transparency. The spectral shape is obtained from \texttt{Komrad}. The probability density for the last scattering as a function of radius, $r$, and comoving propagation direction, $\mu = \cos\theta$, where $\theta$ is the angle between the comoving radial direction and the line-of-sight, is given by equation (B16) in \citet{Beloborodov2011}. The radial cooling profile is obtained from the full scale transfer simulation code \texttt{radshock} \citep{Lundman2018}. The broadened spectrum is then obtained by integrating the local spectrum over radius (or optical depth) and angle. This prescription assumes that there is no structure to the jet, i.e., that the comoving spectrum does not change with angle within $1/\Gamma$ to the line of sight of the observer.

Assume that adiabatic cooling is described by a cooling function $\phi(\tau)$, such that the energy of a photon trapped in its fluid element as measured by a comoving observed at optical depth $\tau$ can be written as 

\begin{equation}\label{eq:AppB_1}
    \epsilon(\tau) = \phi(\tau) \epsilon_{\rm ph}
\end{equation}

\noindent where $\epsilon_{\rm ph}$ is the corresponding photon energy as measured by a comoving observer at the photosphere on the line-of-sight. This prescription neglects the evolution of the spectrum close to the photosphere due to thermalization, i.e., it only accounts for adiabatic cooling. However, we expect the thermalization to be inefficient at small optical depths where the photons are released. 
Indeed, \citet{Beloborodov2011} found that $\sim80$\% of photons are released at optical depths $\tau < 3$. Calculating a Compton $y$-parameter defined as $y = \int_0^3  4\theta_{\rm C, ph} \, \phi(\tau) d\tau$, where $\theta_{\rm C, ph}$ is the value of the Compton temperature at the photosphere and $\phi$ is given by equation \eqref{eq:AppB_5}, we find a mean value for our sample of $\bar{y} \lesssim 0.1$ with the value of $y$ never exceeding 0.5. Thus, we do not expect significant heating over the radii where most photons are released and the low-energy part of the observed spectrum should not be affected by this prescription.

Since all photons are affected similarly by adiabatic cooling, i.e., $\phi$ is not a function of $\epsilon$, the (logarithmic) spectral shape is constant as the photons cool. Specifically, we have that 
\begin{equation}\label{eq:AppB_2}
    \int_{\epsilon_1}^{\epsilon_2}\, N_\epsilon d\epsilon = \int_{\epsilon_{{\rm ph},1}}^{\epsilon_{{\rm ph},2}} N_{\epsilon_{\rm ph}} \, d\epsilon_{\rm ph},
\end{equation}
\noindent where $\epsilon_i = \phi(\tau) \epsilon_{{\rm ph},i}$ as given by equation \eqref{eq:AppB_1}, and $N_\epsilon = dN/d\epsilon$ and $N_{\epsilon_{\rm ph}} = dN_{\rm ph}/d\epsilon_{\rm ph}$ are the photon number spectra at optical depth $\tau$ and the photosphere, respectively. Since this is true for any $\epsilon_1$ and $\epsilon_2$, one gets that $N_\epsilon = N_{\epsilon_{\rm ph}} (\Delta \epsilon_{\rm ph}/\Delta \epsilon) = N_{\epsilon_{\rm ph}}/\phi(\tau)$. 

As photons make their last scattering over a range of radii where the average photon energies are different, the observer receives a spectrum that is broader compared to the comoving photospheric spectrum. Furthermore, the observer receives photons from different scattering angles, which have different Doppler boosts. The observed photon energy is related to the emitted one by $\epsilon_{\rm obs} = D(\mu)\epsilon(\tau) = D(\mu)\phi(\tau)\epsilon_{\rm ph}$, where $D(\mu) = \Gamma(1+\beta\mu) \approx \Gamma(1+\mu)$. As we are in the coasting phase, $\Gamma$ is constant. The observed spectrum will be a superposition of spectra emitted at different optical depths and angles. Let $f(\tau,\mu) d\tau d\mu$ be the probability that an observed photon was emitted at an optical depth between $\tau$ and $\tau+d\tau$ and at an angle between $\mu$ and $\mu + d\mu$. The observed spectrum consisting of these specific photons has the same (logarithmic) spectral shape as the emitted comoving spectrum, just Doppler boosted by a factor $D(\mu)$. Using a similar argument to that of equation \eqref{eq:AppB_2}, one gets $N_{\epsilon_{\rm obs}}\, d\tau d\mu = f(\tau, \mu) d\tau d\mu \, N_{\epsilon_{\rm ph}}/D(\mu)\phi(\tau)$, where $1/D(\mu)\phi(\tau) = \Delta \epsilon_{\rm ph}/\Delta \epsilon_{\rm obs}$. The observed spectrum is obtained by integrating over optical depth and angle as
\begin{equation}\label{eq:AppB_3}
    N_{\epsilon_{\rm obs}} = \iint \frac{N_{\epsilon_{\rm ph}}}{D(\mu)\phi(\tau)} f(\tau, \mu) d\tau d\mu,
\end{equation}
\noindent where $N_{\epsilon_{\rm ph}} = dN_{\rm ph}/{\epsilon_{\rm ph}}$ is everywhere evaluated at $\epsilon_{\rm ph} = \epsilon_{\rm obs} / D(\mu) \phi(\tau)$ (note that $\epsilon_{\rm obs}$ in equation \eqref{eq:AppB_3} is a single, fixed observed photon energy). Thus, it cannot be taken outside of the integrals since it is a function of $\tau$ and $\mu$. By integrating over energy, one can confirm that the prescription conserves particle number:
\begin{equation}\label{eq:AppB_4}
\begin{split}
    N_{\rm obs} &= \int N_{\epsilon_{\rm obs}} \, d\epsilon_{\rm obs} \\
    &= \iiint \frac{N_{\epsilon_{\rm ph}}}{D(\mu)\phi(\tau)} f(\tau, \mu) D(\mu)\phi(\tau) \, d\epsilon_{\rm ph} d\tau d\mu \\
    &= \iiint N_{\epsilon_{\rm ph}} d\epsilon_{\rm ph} \ f(\tau, \mu) d\tau d\mu \\
    &= N_{\rm ph} \iint f(\tau, \mu) d\tau d\mu = N_{\rm ph} \equiv N,
\end{split}
\end{equation}
\noindent where this time, $\int N_{\epsilon_{\rm ph}}d\epsilon_{\rm ph} = N_{\rm ph}$ can be evaluated first and taken outside of the integral, since the total photon number at the photosphere is not a function of $\tau$ or $\mu$. Equation \eqref{eq:AppB_4} also uses the fact that $\iint f(\tau, \mu) d\tau d\mu = 1$. 

Informed by the output of the full radiative transfer simulation code \texttt{radshock}, we find that $\phi(\tau) = C(\tau^{2/3} + 0.2)$ with $C$ being a constant, reproduces the spectrum very well. Using the fact that $\phi(\tau) = 1$ for $\tau = 1$ as required by equation \eqref{eq:AppB_1}, we find that $C = 1/1.2$. Thus,
\begin{equation}\label{eq:AppB_5}
    \phi(\tau) = \frac{\tau^{2/3} + 0.2}{1.2}.
\end{equation}
\noindent This expression is similar to the one found for the first moment of radiation intensity, $I_0$, in \citet{Beloborodov2011}. However, there is a small discrepancy, which may stem from the assumption of isotropy in \citet{Beloborodov2011}.

The expression for how the photon energy evolves with optical depth due to adiabatic cooling given in equation \eqref{eq:AppB_5} is not compatible with the prescription used in \citet{Samuelsson2022}. In \citet{Samuelsson2022}, we numerically stopped the adiabatic cooling in \komrad\ at an optical depth of $\tau = 3$. This was based on Figure 3 in \citet{Beloborodov2011}, where it is shown that the total adiabatic cooling of the photon spectrum as calculated by radiative transfer is equal to an ideal cooling of the photon spectrum (as $\tau^{-2/3})$ if the ideal cooling is stopped at $\tau = 3$. However, in Figure 3 in \citet{Beloborodov2011}, the adiabatic cooling shown is for the luminosity in the lab frame. Within the broadening prescription described in this Appendix, the output from \komrad\ represents the comoving photon distribution at $\tau = 1$. The cooling history (and future cooling for the photons that decouple at $\tau < 1$) of these photons are recreated via equation \eqref{eq:AppB_1}, and integrating over the probability distribution $f(\tau,\mu)d\tau d\mu$ generates the superposition that is the observed spectrum. Thus, it is more accurate to numerically stop the ideal cooling at $\tau = 1.2^{3/2} = 1.31$, as evident from equation \eqref{eq:AppB_5}, which has been incorporated throughout this paper. 

\begin{figure}
    \centering
    \includegraphics[width=\columnwidth]{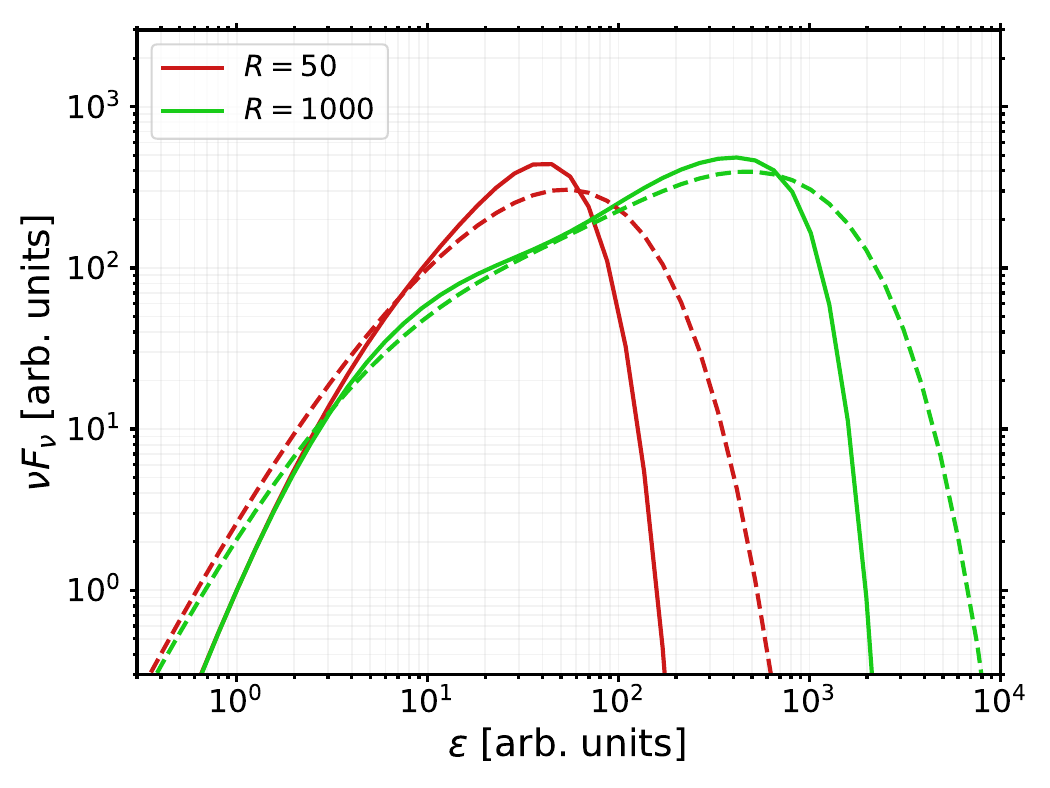}
    \caption{Two spectra before (solid lines) and after (dashed lines) the broadening effects have been added. The solid lines correspond to comoving spectra at the photosphere and the dashed lines to the shapes of the observed spectra, but normalized to have the same average photon energy as the photospheric spectra (see the text for details). The red spectrum has a smaller ratio between the peaks ($R=50$) than the green ($R=1000$). Both initial spectra have been normalized such that $\theta_{u,{\rm K}} = 1$. It is evident that the effect of the broadening scheme is most important for the narrower spectrum. Other parameter values are $\tau\theta = 5$ and $y_r = 1.5$. }
    \label{fig:broadening}
\end{figure}

The probability density function $f(\tau, \mu)$ is found by transforming equation (B16) of \citet{Beloborodov2011} using $\tau = R_\star/r$, which gives
\begin{equation}\label{eq:AppB_6}
\begin{split}
f(\tau, \mu) =& \frac{1}{4}\left\{ \frac{3}{2} + \frac{1}{\pi} \arctan \left[ \frac{1}{3} \left( \tau - \tau^{-1} \right)\right]\right\} \\
&\times \exp \left[-\frac{\tau}{6}\left(3 + \frac{1 - \mu}{1 + \mu}\right)\right].
\end{split}
\end{equation}
The average observed photon energy ${\bar \epsilon}_{\rm obs}$ is given by 
\begin{equation}\label{eq:AppB_7}
\begin{split}
    {\bar \epsilon}_{\rm obs} &= \frac{1}{N} \int \epsilon_{\rm obs} N_{\epsilon_{\rm obs}} \, d\epsilon_{\rm obs} \\
    &= \frac{1}{N} \int \epsilon_{\rm ph} N_{\epsilon_{\rm ph}} d\epsilon_{\rm ph} \iint D(\mu)\phi(\tau) f(\tau, \mu) d\tau d\mu \\
    &= \Gamma {\bar \epsilon}_{\rm ph} \iint (1+\mu) \phi(\tau) f(\tau, \mu) d\tau d\mu.
\end{split}
\end{equation}
\noindent With $\phi(\tau)$ and $f(\tau, \mu)$ given by equations \eqref{eq:AppB_5} and \eqref{eq:AppB_6}, we find that the double integral in the last line of equation \eqref{eq:AppB_7} equals 5/3. Thus, the average photon energy in the observed spectrum (not accounting for redshift) is $5\Gamma/3$ times larger compared to ${\bar \epsilon}_{\rm ph}$. 

Since the broadening is linear with $\Gamma$, we can model it as a post-processing parameter. We normalize the output broadened spectrum such that the average photon energy in the broadened spectrum equals ${\bar \epsilon}_{\rm ph}$. Thus, to obtain the observed spectrum, the energy grid of the broadened spectrum should be multiplied by $5\Gamma/3$, which motivates the choice of the Doppler factor used in Section \ref{sec:Results}.
In Figure \ref{fig:broadening}, we show the effect of the broadening scheme on two spectra with different values of $R$. The effect of the broadening is more prominent for the narrower spectrum.

\section{Additional configurations}\label{App:different_params}

\begin{table}
    \centering
    \caption{Lower and upper limits for the uniform distributions used to generate the histograms shown in Figure \ref{fig:different_params_flat}}
    \begin{tabular}{ccc}
    \hline
         Parameter & Lower limit & Upper limit  \\
    \hline
         $\Gamma_s$ & 5 & 200 \\
         $\tau_s$ & 5 & 300 \\
         $\psi$ & 2 & 25 \\
    \hline
    \end{tabular}
    \label{tab:different_params}
\end{table}

\begin{table}[t]
    \centering
    \caption{Mean and standard deviations for the normal distributions used to generate the histograms shown in Figure \ref{fig:different_params_high_G}. The logarithms used for $\Gamma_s$ and $\tau_s$ are base 10}
    \begin{tabular}{ccc}
    \hline
         Parameter & Mean & Standard deviation  \\
    \hline
         $\log(\Gamma_s)$ & 2 & 0.2 \\
         $\log(\tau_s)$ & 1.5 & 0.3 \\
         $\psi$ & 15 & 5 \\
    \hline
    \end{tabular}
    \label{tab:different_params2}
\end{table}

\begin{figure*}
    \centering
    \includegraphics[width=0.75\columnwidth]{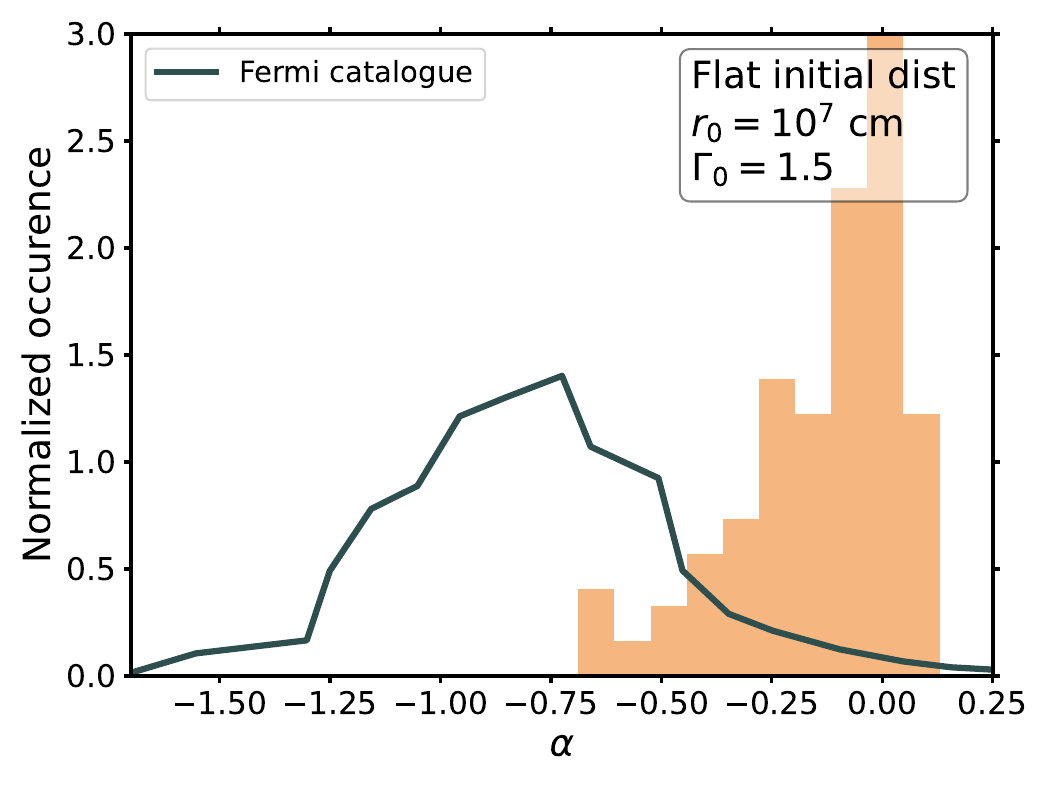}
    \includegraphics[width=0.75\columnwidth]{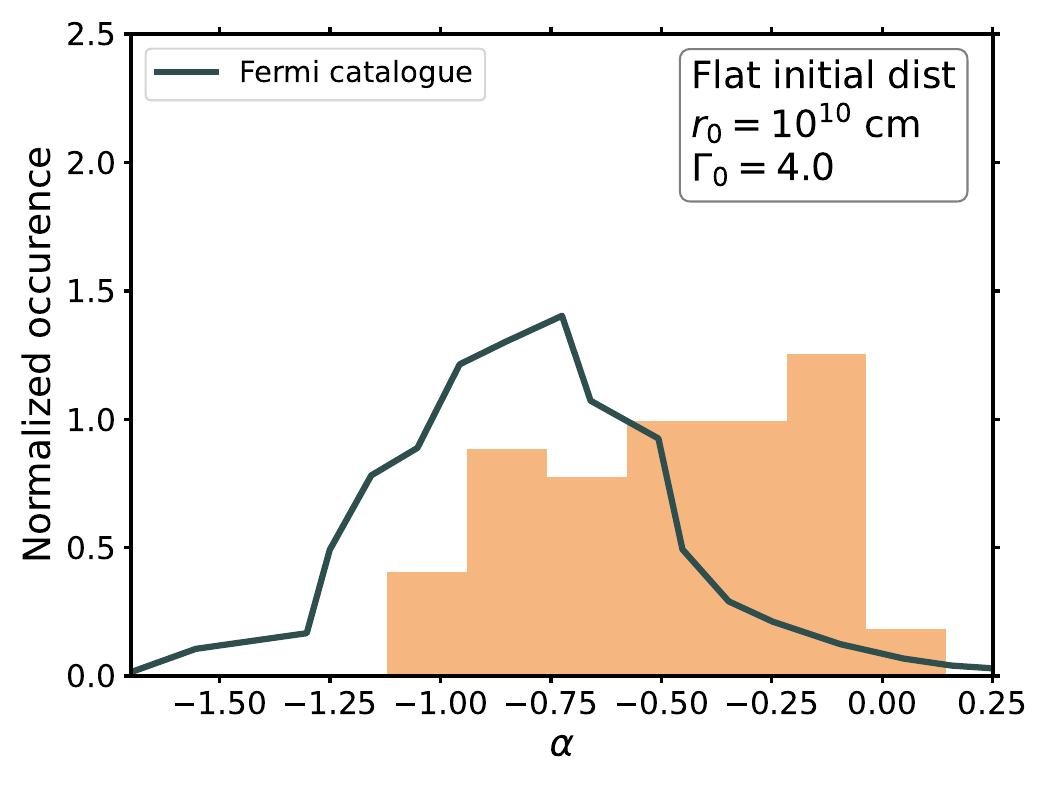}\\
    \includegraphics[width=0.75\columnwidth]{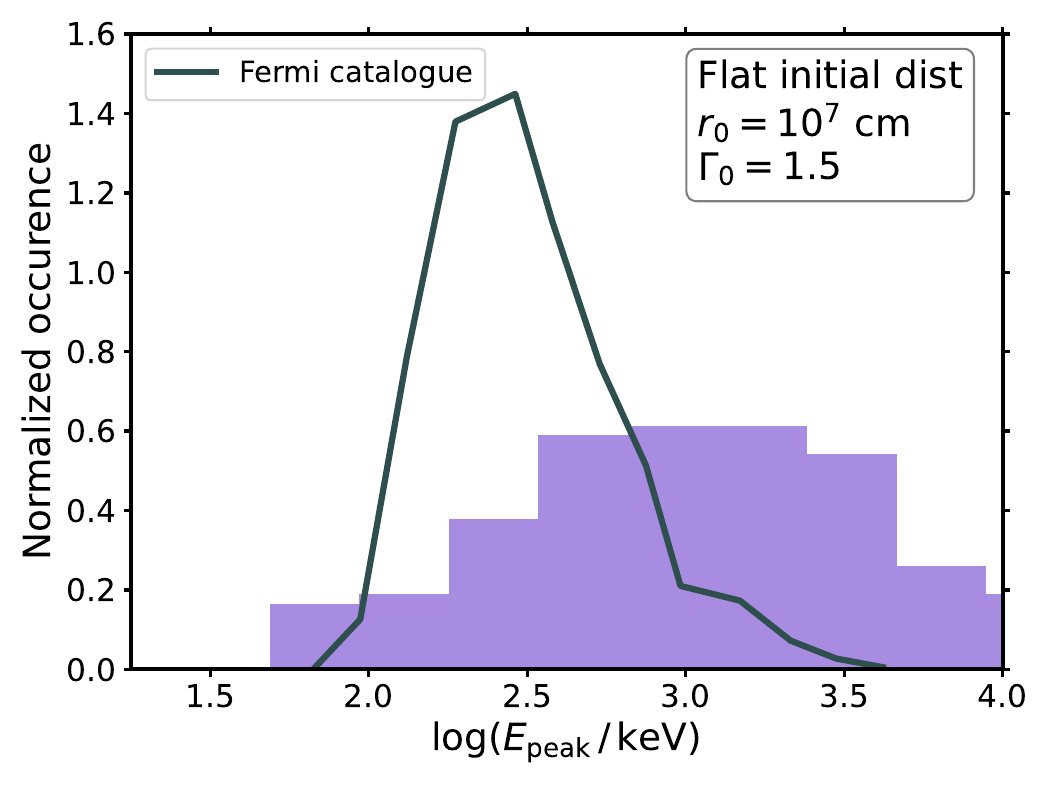}
    \includegraphics[width=0.75\columnwidth]{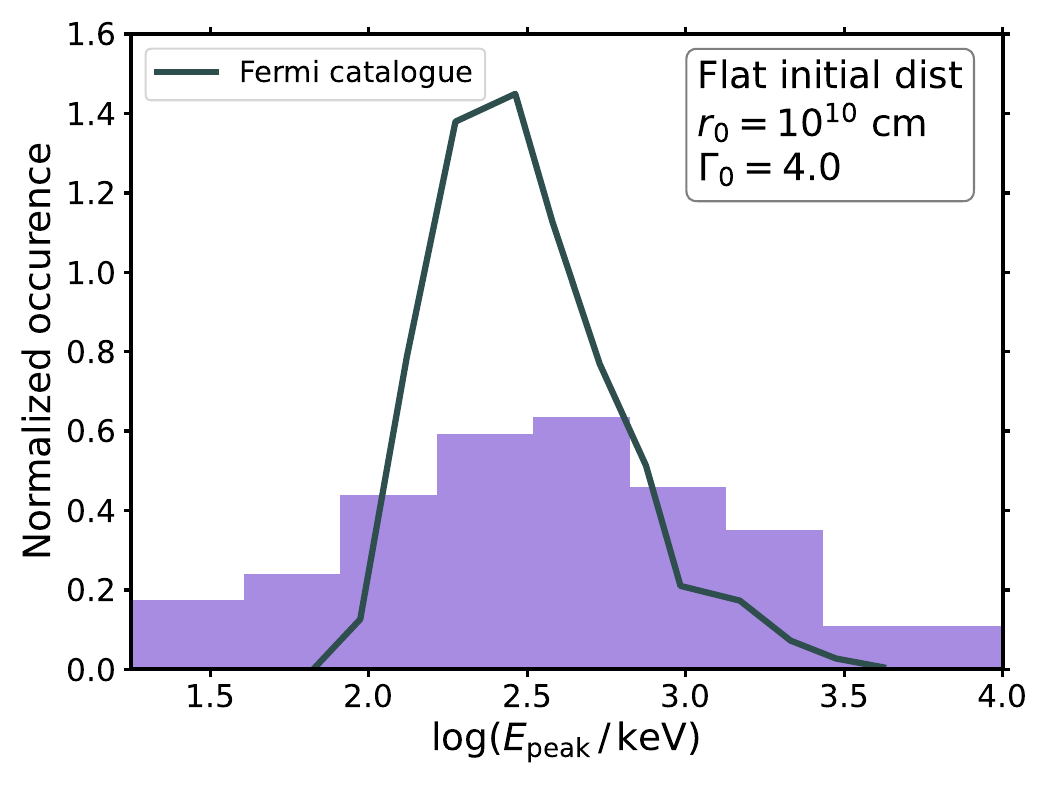}
    \caption{Similar to Figure \ref{fig:hist_CPL}, but generated from the higher order initial parameter distribution given in Table \ref{tab:different_params}. This choice of initial parameters results in higher values of $\tau_i$ on average, leading to harder spectra. Many of these burst would likely be missed in observations due to the increase in adiabatic degrading and decrease in luminosity.}
    \label{fig:different_params_flat}
\end{figure*}

\begin{figure*}
    \centering
    \includegraphics[width=0.75\columnwidth]{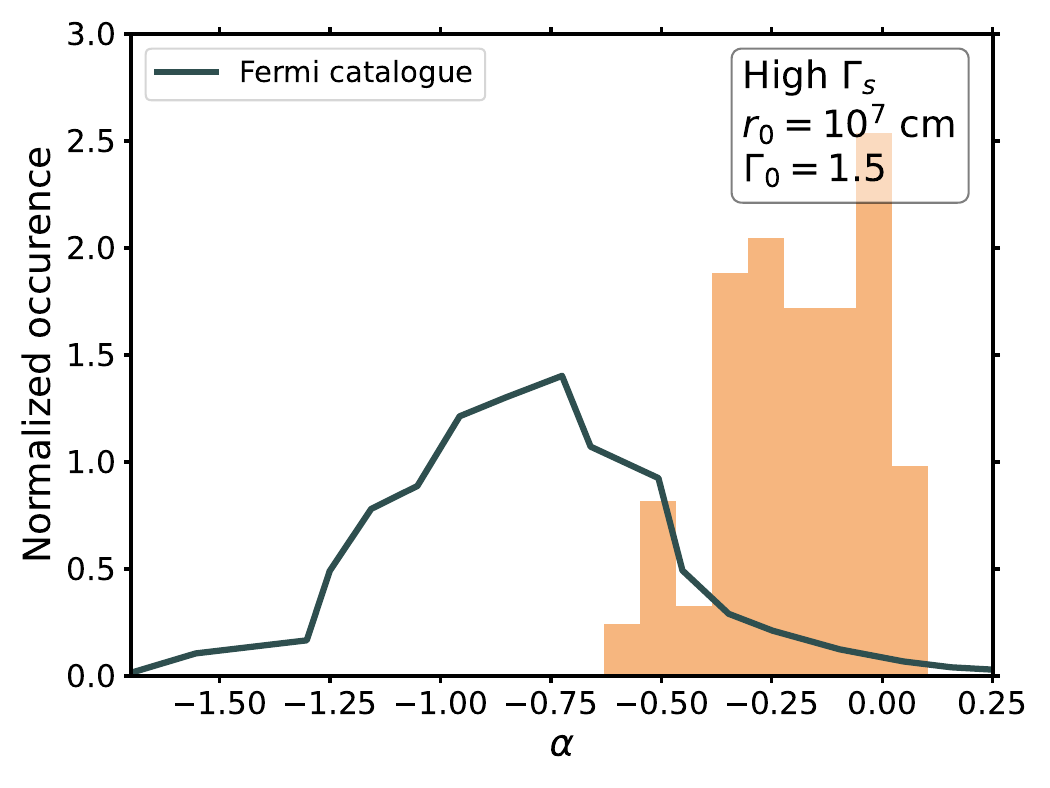}
    \includegraphics[width=0.75\columnwidth]{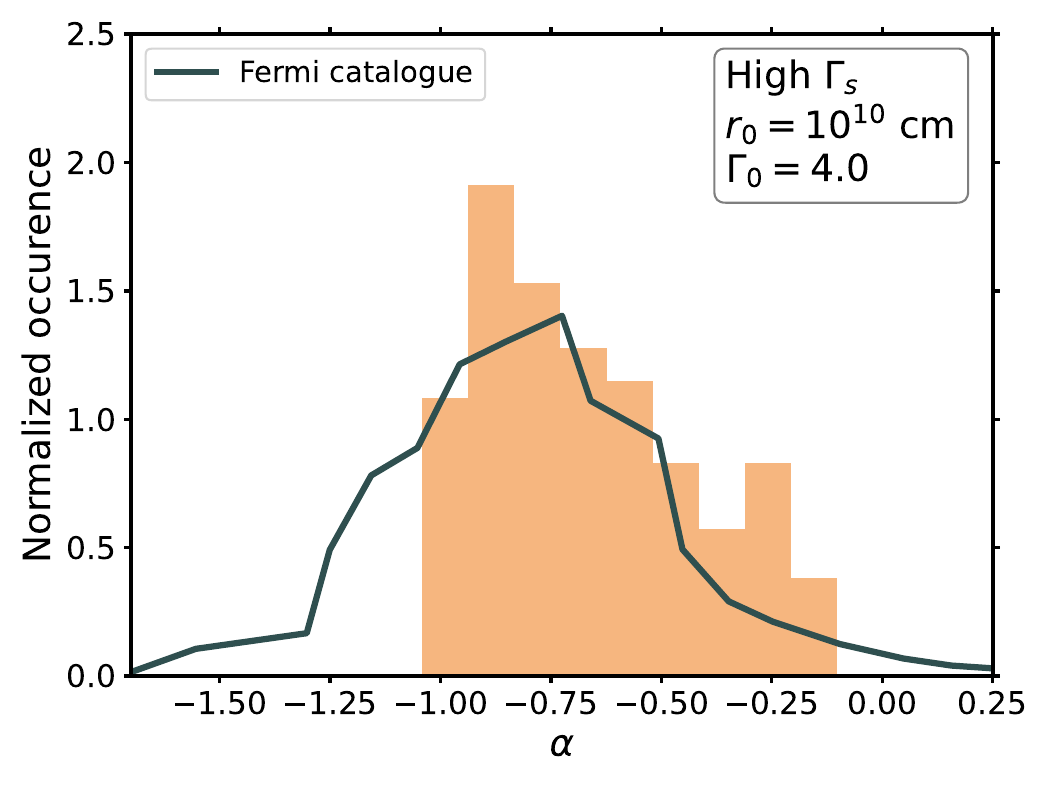}\\
    \includegraphics[width=0.75\columnwidth]{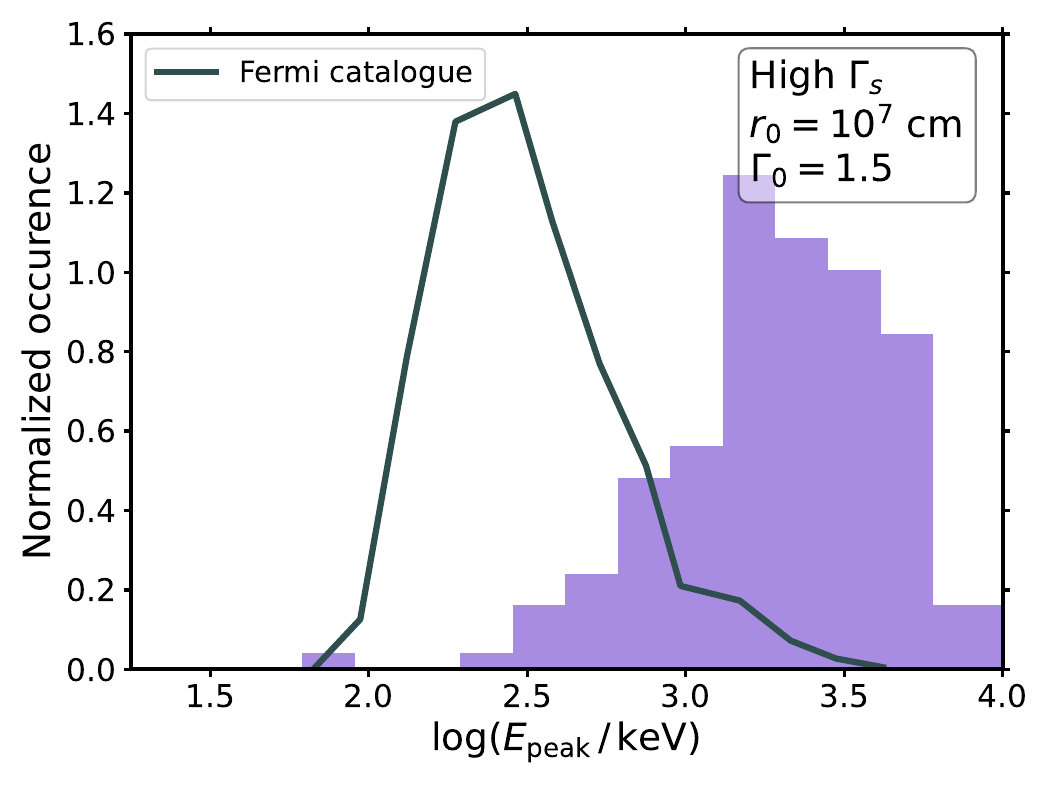}
    \includegraphics[width=0.75\columnwidth]{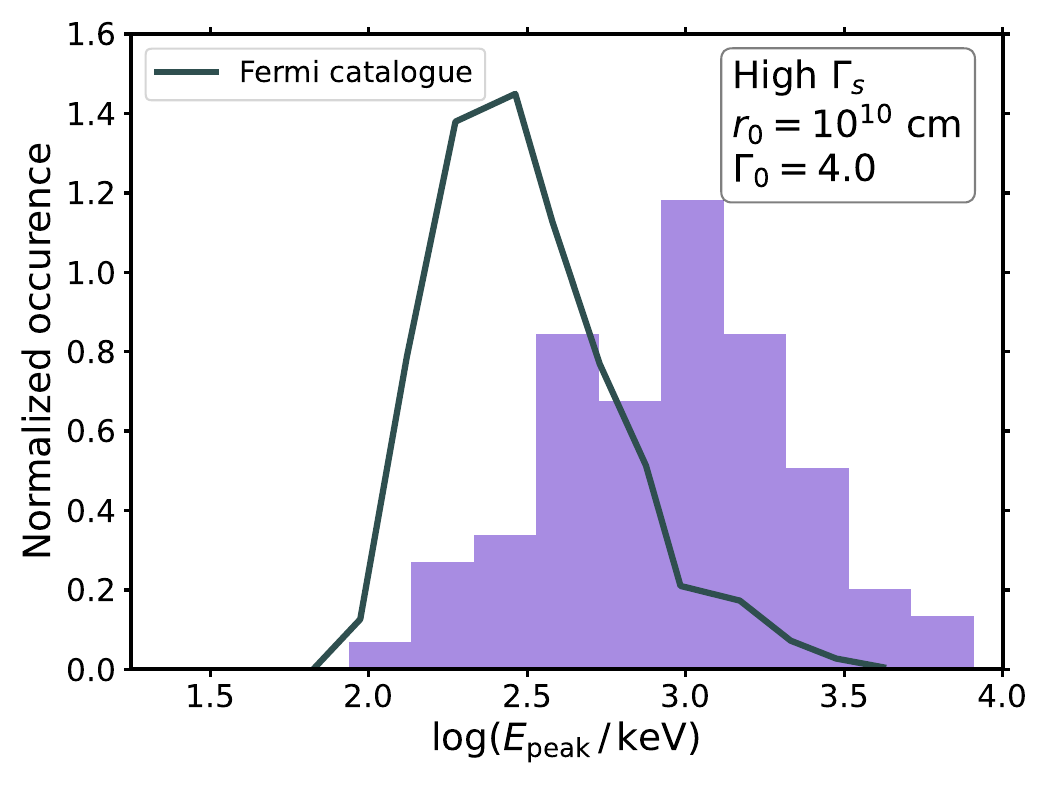}
    \caption{Similar to Figure \ref{fig:hist_CPL}, but generated from the higher order initial parameter distribution given on the in Table \ref{tab:different_params2}. A higher initial value of $\Gamma_s$ has a small effect on $\alpha$ but increases the observed peak energy.}
    \label{fig:different_params_high_G}
\end{figure*}

\begin{figure*}
    \centering
    \includegraphics[width=0.75\columnwidth]{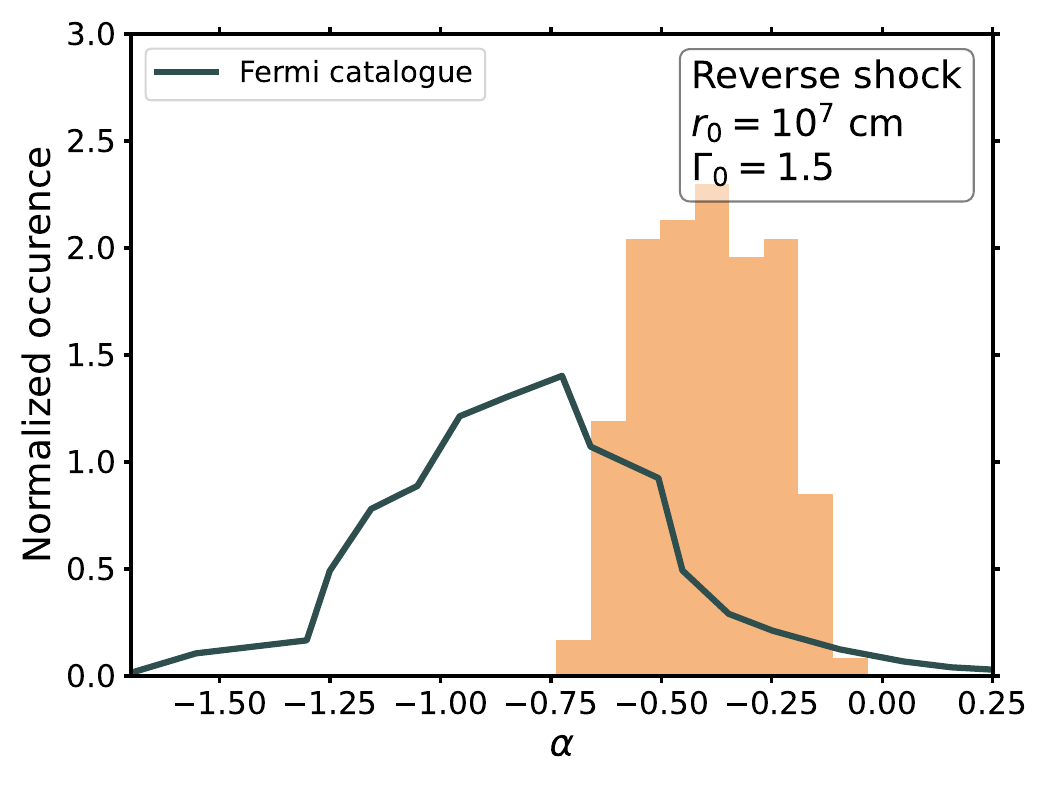}
    \includegraphics[width=0.75\columnwidth]{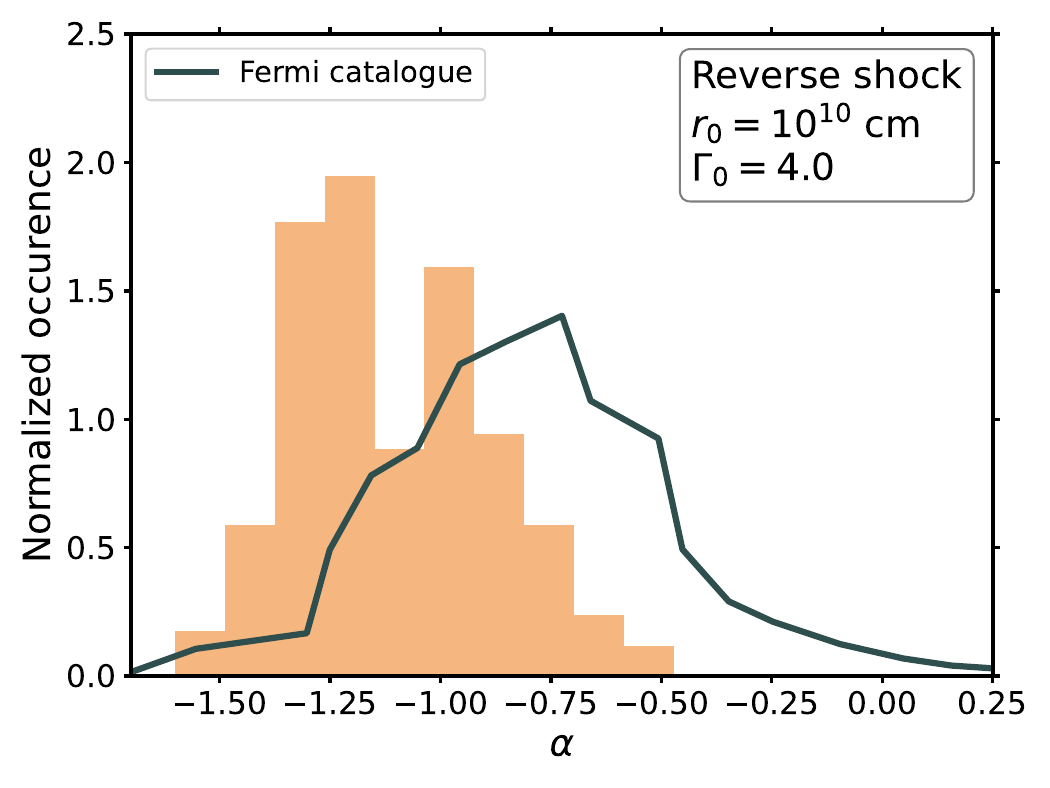}\\
    \includegraphics[width=0.75\columnwidth]{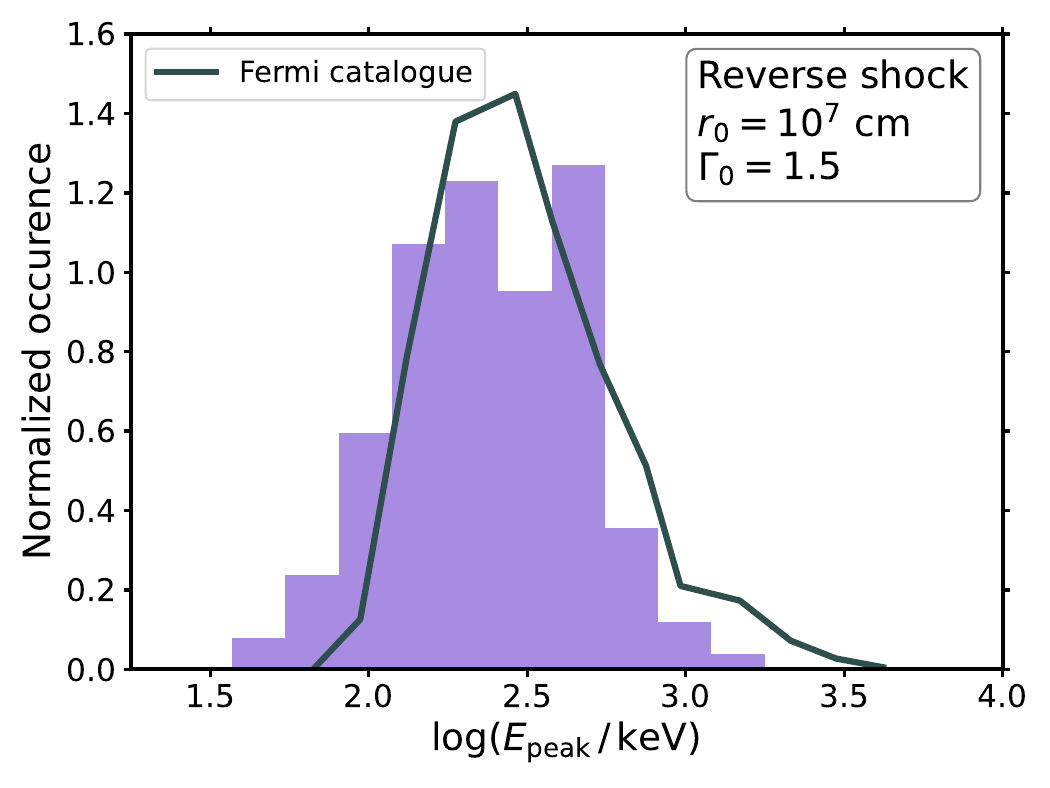}
    \includegraphics[width=0.75\columnwidth]{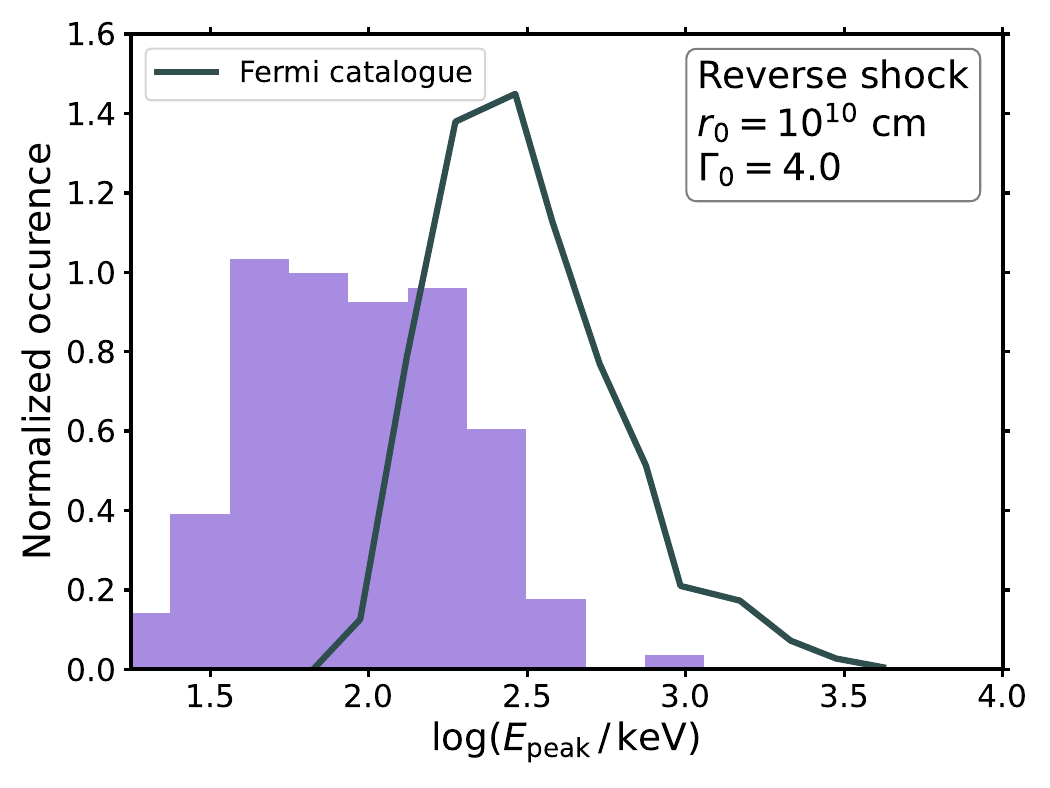}
    \caption{Similar to Figure \ref{fig:hist_CPL}, but using the conditions for the reverse shock instead of the forward shock in equations \eqref{eq:theta_u2} and \eqref{eq:tilde_n}. The resulting spectra has lower values of $E_{\rm peak}$ and are softer, with $\alpha$ reaching $\sim -1.5$ in a few cases in the delayed acceleration scenario. }
    \label{fig:different_params_reverse_shock}
\end{figure*}

\begin{figure*}
    \centering
    \includegraphics[width=0.65\columnwidth]{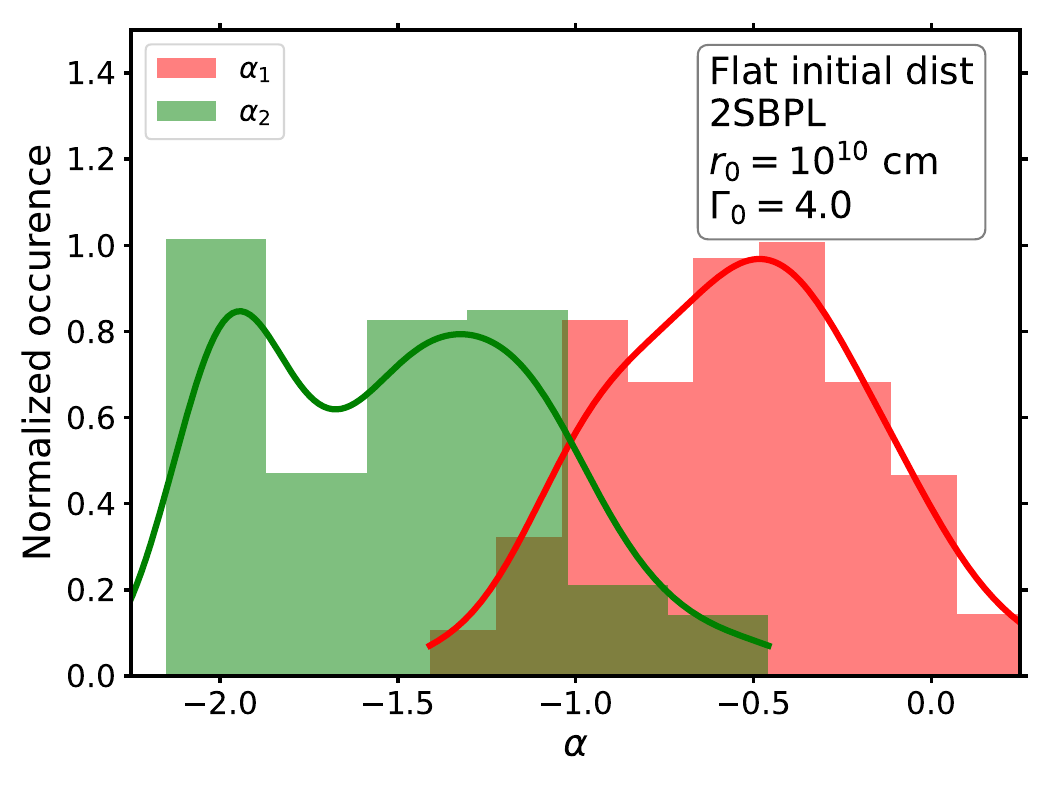}
    \includegraphics[width=0.65\columnwidth]{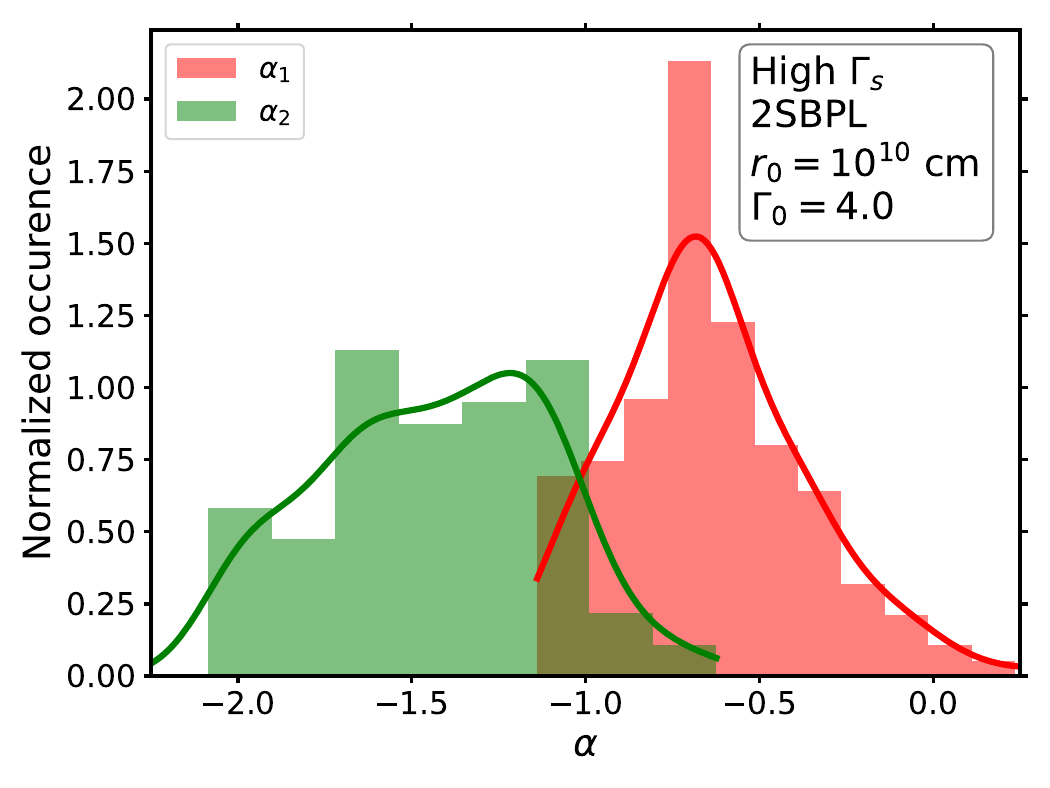}
    \includegraphics[width=0.65\columnwidth]{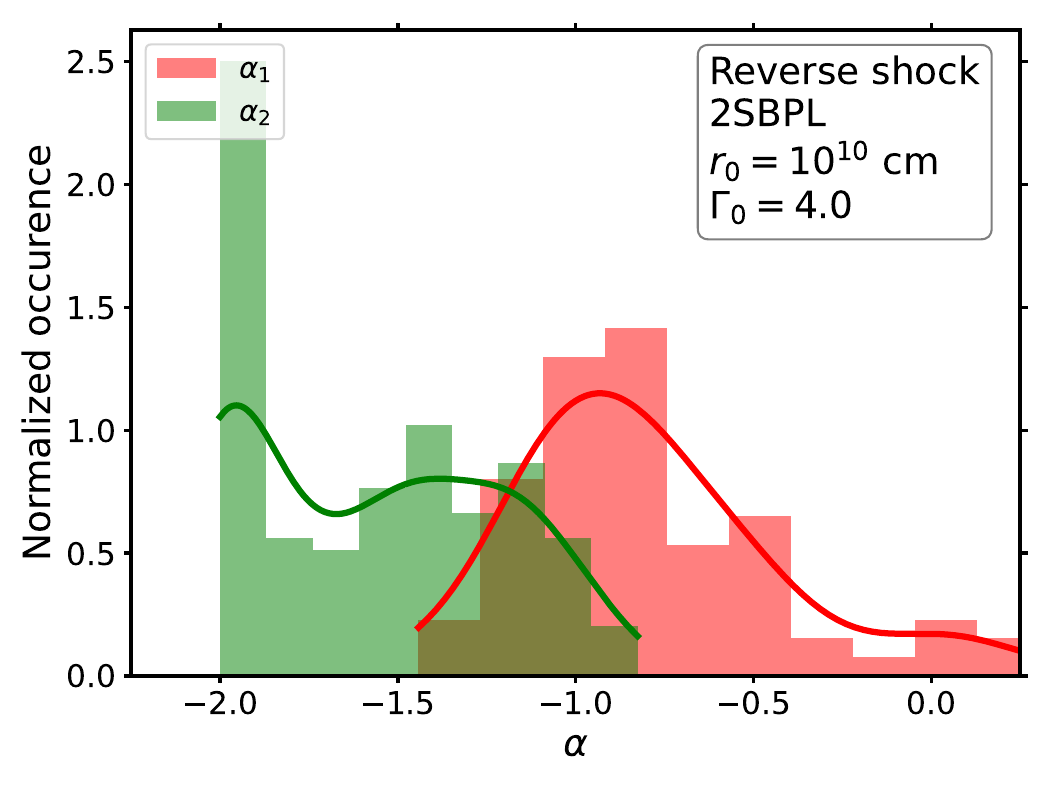}\\
    \includegraphics[width=0.65\columnwidth]{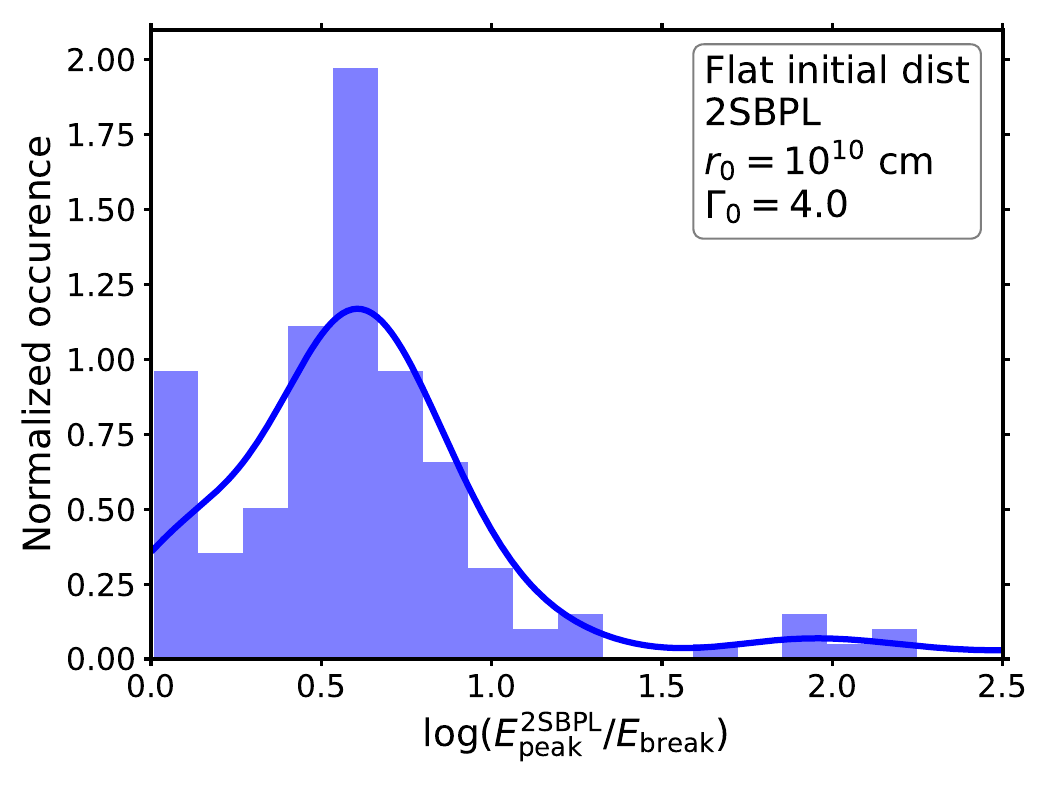}
    \includegraphics[width=0.65\columnwidth]{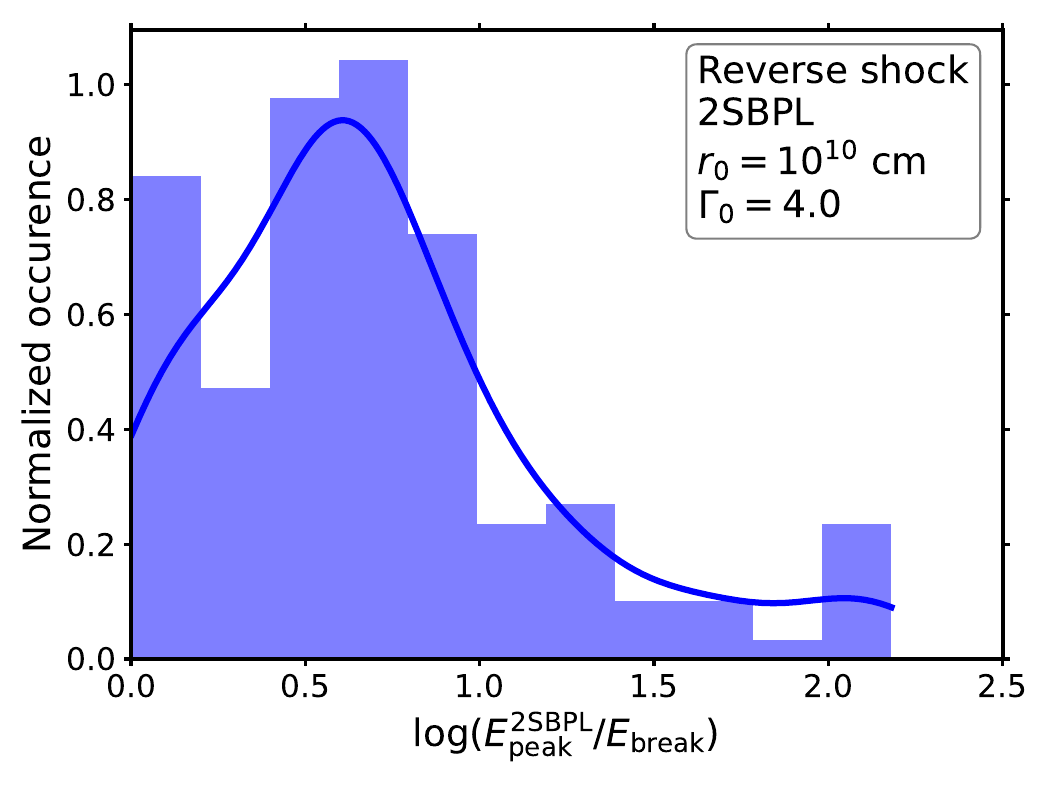}
    \includegraphics[width=0.65\columnwidth]{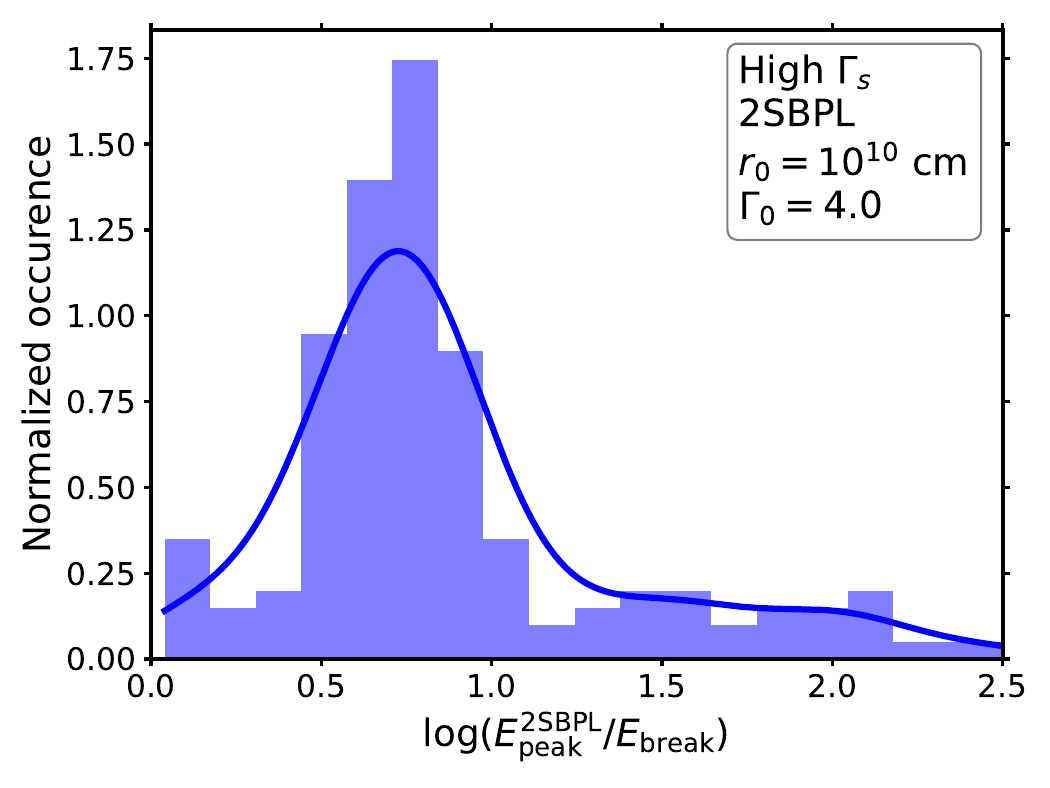}
    \caption{Similar to Figure \ref{fig:hist_2SBPL}, but for the conditions used to generate Figure \ref{fig:different_params_flat} (left), Figure \ref{fig:different_params_high_G} (middle), and Figure \ref{fig:different_params_reverse_shock} (right). The discussion in subsection \ref{sec:disc_2SBPL} applies in these cases as well. In the middle panels where $\Gamma_s$ is increased, the curvature of the spectrum is inside the detector window and the excess of bursts with $\alpha_2 \sim -2$ vanishes. Furthermore, the spectral indices are more clustered around the values expected in marginally fast cooling synchrotron emission.}
    \label{fig:different_2SBPL}
\end{figure*}

In this appendix, we present results similar to those shown in the main text for three additional configurations of the internal shock parameters. The first two (Figures \ref{fig:different_params_flat} and \ref{fig:different_params_high_G}) have different distributions for the higher-order initial parameters. For Figure \ref{fig:different_params_flat}, uniform distributions have been used, as given in Table \ref{tab:different_params}. In Figure \ref{fig:different_params_high_G}, the distributions are similar to the main text except that the average value of $\Gamma_s$ has been increased, as given in Table \ref{tab:different_params2}. The methodology is the same as when generating the histograms presented in Figure \ref{fig:hist_CPL}. Specifically, the cuts outlined in subsection \ref{sec:implementation} still apply. This leads to a removal of $\sim 30$\% of the generated initial parameter triplets in the case of Figure \ref{fig:different_params_flat}, mostly due to many parameter combinations resulting in $\tau\theta > 50$.
For the third additional configuration (Figure \ref{fig:different_params_reverse_shock}), the distributions for the higher-order initial parameters is given in Table \ref{tab:init_dist}, i.e., the same as for the results presented in the main text. However, it uses the upstream temperature and photon-to-proton ratio obtained for the reverse shock instead of the forward shock in equations \eqref{eq:theta_u2} and \eqref{eq:tilde_n}. This leads to softer spectra on average, with values for $\alpha$ reaching as low as $-1.5$ in the delayed acceleration scenario. In Figure \ref{fig:different_2SBPL}, we show the obtained 2SBPL parameter distributions for the three additional configurations, similar to Figure \ref{fig:hist_2SBPL}.

\end{document}